\documentclass[10pt,apjl]{emulateapj}
\usepackage[dvips]{epsfig}
\usepackage[dvips]{color}

\usepackage{apjfonts}

\begin{document}
\newcommand{\hes} {HE~0557$-$4840}
\newcommand{\hen} {HE~0107$-$5240}
\newcommand{\hea} {HE~1327$-$2326}
\newcommand{\cd}  {CD$-38^{\circ}\,245$}
\newcommand{\kms} {\rm km s$^{-1}$} 
\newcommand{\vradgeo} {$v_{\mbox{\scriptsize rad,geo}}$}
\newcommand{\vradhelio} {$v_{\mbox{\scriptsize rad,helio}}$}
\newcommand{\teff}  {$T_{\rm eff}$} 
\newcommand{\logg}  {$\log g$} 
\newcommand{\loggf} {log~$gf$} 
\newcommand{\bvz}   {$(B-V)_{0}$} 
\newcommand{\vhes}  {$V_{\rm HES}$} 
\newcommand{\bmag}  {$B$} 
\newcommand{\bvhes} {$(B-V)_{\rm HES}$} 
\newcommand{\fehk}  {[Fe/H]$_{\rm K}$}
\newcommand{\kp}    {K$^{\prime}$}
\newcommand{\gp}    {G$^{\prime}$}
\newcommand{\hp}    {H$^{\prime}$}

\shorttitle{THE MOST METAL-POOR STARS I.}
\shortauthors{Norris et al.}

\title{THE MOST METAL-POOR STARS. I. DISCOVERY, DATA, AND ATMOSPHERIC
  PARAMETERS\altaffilmark{1}}

\altaffiltext{1}{This paper includes data obtained with the ANU 2.3\,m
  Telescope at Siding Spring Observatory, Australia; the Magellan Clay
  Telescope at Las Campanas Observatory, Chile; the Keck I Telescope
  at the W.\ M.\ Keck Observatory, Hawaii, USA; and the VLT (Kueyen)
  of the European Southern Observatory, Paranal, Chile (proposal
  281.D-5015).}

\author{
JOHN E. NORRIS\altaffilmark{2},
M. S. BESSELL\altaffilmark{2},
DAVID YONG\altaffilmark{2},
N. CHRISTLIEB\altaffilmark{3},
P. S. BARKLEM\altaffilmark{4},
M. ASPLUND\altaffilmark{2,5},
SIMON J. MURPHY\altaffilmark{2},
TIMOTHY C. BEERS\altaffilmark{6,7},
ANNA FREBEL\altaffilmark{8}, AND 
S. G. RYAN\altaffilmark{9}}

\altaffiltext{2}{Research School of Astronomy and Astrophysics, The
  Australian National University, Weston, ACT 2611, Australia;
  jen@mso.anu.edu.au, bessell@mso.anu.edu.au, yong@mso.anu.edu.au,
  martin@mso.anu.edu.au}

\altaffiltext{3}{Zentrum f\"ur Astronomie der Universit\"at
  Heidelberg, Landessternwarte, K{\"o}nigstuhl 12, D-69117 Heidelberg,
  Germany; n.christlieb@lsw.uni-heidelberg.de}

\altaffiltext{4}{Department of Physics and Astronomy, Uppsala
University, Box 515, 75120 Uppsala, Sweden;
paul.barklem@physics.uu.se}

\altaffiltext{5}{Max Planck Institute for Astrophysics, Postfach 1317, 
  85741 Garching, Germany}

\altaffiltext{6}{National Optical Astronomy Observatory, Tucson, AZ
85719, USA}

\altaffiltext{7}{Department of Physics \& Astronomy and JINA: Joint
  Institute for Nuclear Astrophysics, Michigan State University,
  E. Lansing, MI 48824, USA; beers@pa.msu.edu}

\altaffiltext{8}{Department of Physics, Massachusetts Institute of
  Technology, Cambridge, MA 02139, USA; afrebel@mit.edu}

\altaffiltext{9}{Centre for Astrophysics Research, School of Physics,
  Astronomy \& Mathematics, University of Hertfordshire, College Lane,
  Hatfield, Hertfordshire, AL10 9AB, UK; s.g.ryan@herts.ac.uk}

\begin{abstract}

We report the discovery of 34 stars in the Hamburg/ESO Survey for metal-poor stars and the Sloan Digital Sky Survey that have [Fe/H] $\la$ --3.0.  Their median and minimum abundances are [Fe/H] = --3.1 and --4.1, respectively, while 10 stars have [Fe/H] $<$ --3.5. High-resolution, high-$S/N$ spectroscopic data -- equivalent widths and radial velocities -- are presented for these stars, together with an additional four objects previously reported or currently being investigated elsewhere.  We have determined the atmospheric parameters, effective temperature (\teff) and surface gravity (\logg), which are critical in the determination of the chemical abundances and the evolutionary status of these stars.  Three techniques were used to derive these parameters.  Spectrophotometric fits to model atmosphere fluxes were used to derive {\teff}, {\logg}, and an estimate of $E(B-V)$; H$\alpha$, H$\beta$, and H$\gamma$ profile fitting to model atmosphere results provided the second determination of {\teff} and {\logg}; and finally, we used an empirical \teff\--calibrated H$\delta$ index, for the third, independent \teff\ determination. The three values of \teff\ are in good agreement, although the profile fitting may yield systematically cooler \teff\ values, by $\sim100$K.  This collective data set will be analyzed in future papers in the present series to utilize the most metal-poor stars as probes of conditions in the early Universe.

\end{abstract}

\keywords{Cosmology: Early Universe, Galaxy: Formation, Galaxy: Halo,
  Stars: Abundances, Stars: Fundamental Parameters}

\section{INTRODUCTION} \label{Sec:intro}

Six decades after the discovery of metal-poor stars by
\citet{chamberlain51}, the study of these objects has become a mature
area of research.  In 2010, high-resolution, high-signal-to-noise
($S/N$) chemical abundance analyses existed for some 400 stars that
have [Fe/H] $< -2.5$, some 24 with [Fe/H] $< -3.5$, and three having
[Fe/H] $< - 4.5$ (see the compilation of
\citet{frebel10}\footnote{https://www.cfa.harvard.edu/$\sim$afrebel/}) .
As discussed by many authors, the most metal-poor stars, believed to
have formed at redshifts z $\ga$~6, are among the best probes of
conditions in the early Universe, including in particular the
formation of the first stars and the first chemical elements.  We
shall not repeat here the case for this endeavor, but refer the
reader to earlier works (see, e.g., \citealt{bessell84, mcwilliam95,
  ryanetal96, norris01, cayrel04, beers05, cohen08, lai08, frebel11})
for a thorough discussion of the rationale that drives this very
active field.

The aim of the present series of papers is to increase the inventory
of the most metal-poor stars, by which we mean [Fe/H] $\la -3.5$, in
order to understand the Metallicity Distribution Function (MDF) at
lowest metallicity and to investigate abundance patterns that contain
clues to conditions at the earliest times.  We are interested to gain
insight into the relative abundances ([X/Fe]\footnote{[X/Fe] =
  $\log_{10}(\rm N_{\rm X}/N_{\rm Fe})_\star - \log_{10}(N_{\rm
    X}/N_{\rm Fe})_\odot$}) in this abundance regime, involving in
particular C, N, O, Mg, Na, Al, and the heavy neutron-capture
elements, which deviate strongly from the relatively well-defined
behavior of the majority of stars in the ([X/Fe], [Fe/H])-- planes at
higher abundances ([Fe/H] $>$ --3.5) (see, e.g.,
\citealt[Figure~11]{norris07}).

Section 2 of the paper is concerned with the discovery of stars having
[Fe/H] $< -3.0$, by obtaining medium-resolution spectra of candidate
metal-poor stars.  Section 3 describes follow-up high-resolution,
high-$S/N$ spectroscopy of the most metal-poor of them, together with
the reduction and analysis of the spectra to produce the basic data
(equivalent widths and radial velocities) upon which subsequent
analyses will be based.  Our sample contains 38 stars, some 13 of
which (12 identified in the endeavors described in Section 2) have
[Fe/H] $< -3.5$, based on subsequent high-resolution, high-$S/N$
spectroscopic analysis.  In Section IV we describe the determination
of accurate effective temperatures needed for the chemical abundance
analysis.  Papers II, III (Yong et al.\, 2012a, b) and Papers IV, V
(Norris et al.\, 2012, b) will present accurate chemical abundances and
discussion of the elements from lithium through to the
heavy-neutron-capture elements.

\newpage

\section{DISCOVERY AND SELECTION OF PROGRAM STARS} \label{Sec:discovery}

The present program represents the completion of a search over some 30
years for the most metal-poor stars undertaken at the Mount Stromlo \&
Siding Spring Observatories, ANU (now known as RSAA, ANU), based on
techniques involving (in the main part) high-proper-motion stars (the
NLTT survey (\citealt{luyten79, luyten80})) and metal-weak candidates
obtained from Schmidt wide-field objective-prism surveys (the HK
Survey \citep {beers85, beers92}) and the Hamburg/ESO Survey (HES
\citep{christlieb08}).

Results based at least in part on these efforts have already been
presented by \citet{aoki02, aoki06}, \citet{bessell84},
\citet{cayrel04}, \citet{christlieb02, christlieb04}, \citet{frebel05,
  frebel06, frebel07a, frebel07b}, \citet{honda04}, \citet{li10},
\citet{norris85,norris99,norris01,norris07}, \citet{ryan91},
\citet{ryanetal91,ryanetal96}, and \citet{schoerck09}.  These include
the discovery and analysis of three of the four most metal-poor stars
currently known\footnote{For details of the fourth star,
  SDSS~J102915+172927, see \citet{caffau11}.}: {\hen} with [Fe/H] =
--5.3 (\citealt{christlieb02, christlieb04}), {\hea} with [Fe/H] =
--5.4 (\citealt{frebel05, aoki06}), and {\hes} with [Fe/H] = --4.8
\citep{norris07}.

The current sample of the most metal-poor stars is based principally on a
medium-resolution spectroscopic survey of metal-poor candidates from the
HES, supplemented by a few stars known to be extremely metal-poor from
other sources.

\subsection{Medium-resolution Spectroscopy with the ANU 2.3\,m Telescope} \label{Sec:2.3m}

Metal-poor candidates from the Hamburg/ESO objective-prism survey have
been observed during the present investigation with the Australian
National University's 2.3\,m Telescope/Double Beam Spectrograph
combination on Siding Spring Mountain, during observing sessions in
2005--2009.  The spectra have a resolving power R $\sim$ 1600, and
cover the wavelength range 3600--5400\,{\AA}.  They were reduced with
the FIGARO package\footnote{http://www.aao.gov.au/figaro} and
flatfielded and wavelength calibrated by using spectra of quartz and
Fe-Ar lamps, respectively.  Following \citet[their Table~2]{beers99},
we measured the CaII K line index, {\kp}, the CH G-band index, {\gp},
and the hydrogen indices H$\gamma$ and H$\delta$ to form {\hp}, the
mean of the two estimates.  ({\hp} was not measured for objects with
{\gp} $>$ 4.0\,{\AA}, for which the index is affected by strong CH
absorption).  We used these data, following the precepts of Beers et
al., to obtain estimates of iron abundance, here designated {\fehk}.
Analysis of some of these spectra have also been used for discussion
of the MDF of the Galactic halo by \citet{schoerck09} and
\citet{li10}, to which we refer the reader.  For the present work, we
recall that we were interested in discovering stars with [Fe/H] $\la$
--3.0, and have adopted techniques that differ in two respects from
those works.  First, our investigation used the original abundance
calibration of \citet{beers99}.  Second, that technique requires
values not only of {\kp} but also of {\bvz}.  For {\bvz} $\la$ 0.70 we
used the hydrogen index, when available, to provide the color
estimate\footnote{Based on colors and reddenings of high-proper-motion
  stars presented by \citet{carney94}, together with values of {\kp}
  we obtained with the equipment described above as part of another
  investigation.}: {\bvz} = 0.840 -- 0.1541{\hp} +
0.01148{\hp}{$^{2}$}; otherwise, we used {\bvhes} from the HES survey
(see \citealt{christlieb08}), corrected for reddening following
\citet{schlegel98}.

In what follows we present data for stars that were discovered to have
{\fehk} $\la$~--3.0, based on the above medium-resolution spectroscopy
obtained in 2005--2008.  Results for some extremely metal-poor stars
that were observed in 2005 (e.g., {\hes}, [Fe/H] = --4.8) have already
been published \citep{norris07} or are the subject of analysis
currently underway \citep{garcia08}.  Followup investigation of
objects observed in 2009 is work for the future.

During 2005-2009 we obtained some 3400 spectra of HES metal-poor
candidates with the 2.3\,m Telescope, and from those observed in
2005-2008 selected the 1460 stars that satisfy the following criteria:
(1) B magnitudes in the range 12.6 $<$ B$_{\rm HES}$ $<$ 16.5; (2)
{\hp}, the mean of the \citet{beers99} hydrogen H$\gamma$ and
H$\delta$ indices, less than 5.5\,{\AA} (i.e. {\bvz} $\ga$~0.34) and
{\bvhes} $\ga$ 0.30; (3) photon counts at 4100\,{\AA} greater than 200
per $\sim$1\,{\AA} pixel; and (4) spectra exhibit no anomalies such as
hydrogen or Ca~II emission.  The first criterion was chosen to
discover metal-poor stars that could be observed with 6--10\,m class
telescopes to obtain spectra at high resolution and high-$S/N$ in
times less than a few hours; the second confines the selection almost
exclusively to main-sequence dwarfs, and to red giant branch (RGB) and
red horizontal branch (RHB) stars; and the third accepts
medium-resolution spectra with sufficient $S/N$ to produce abundance
accuracy of $\Delta$[Fe/H]~$\sim$~0.2--0.3~dex -- at least for stars
that are not carbon-rich (and for which abundances from
medium-resolution spectroscopy are less well-determined).  For the 24 HES
dwarfs and giants in the present sample with the CH G-band index {\gp}
$<$ 1.5\,{\AA}, the dispersion of the differences between the present
values of {\fehk} and those we obtain from the analysis of our
high-resolution, high-$S/N$, spectra is 0.30 dex, while for the
remaining seven (carbon-rich) objects, which have {\gp} $>$
3.5\,{\AA}, the dispersion of the abundance differences is 0.40 dex.

\begin{figure}[!tbp]
\figurenum{1}
\begin{center}
\includegraphics[width=8.5cm,angle=0]{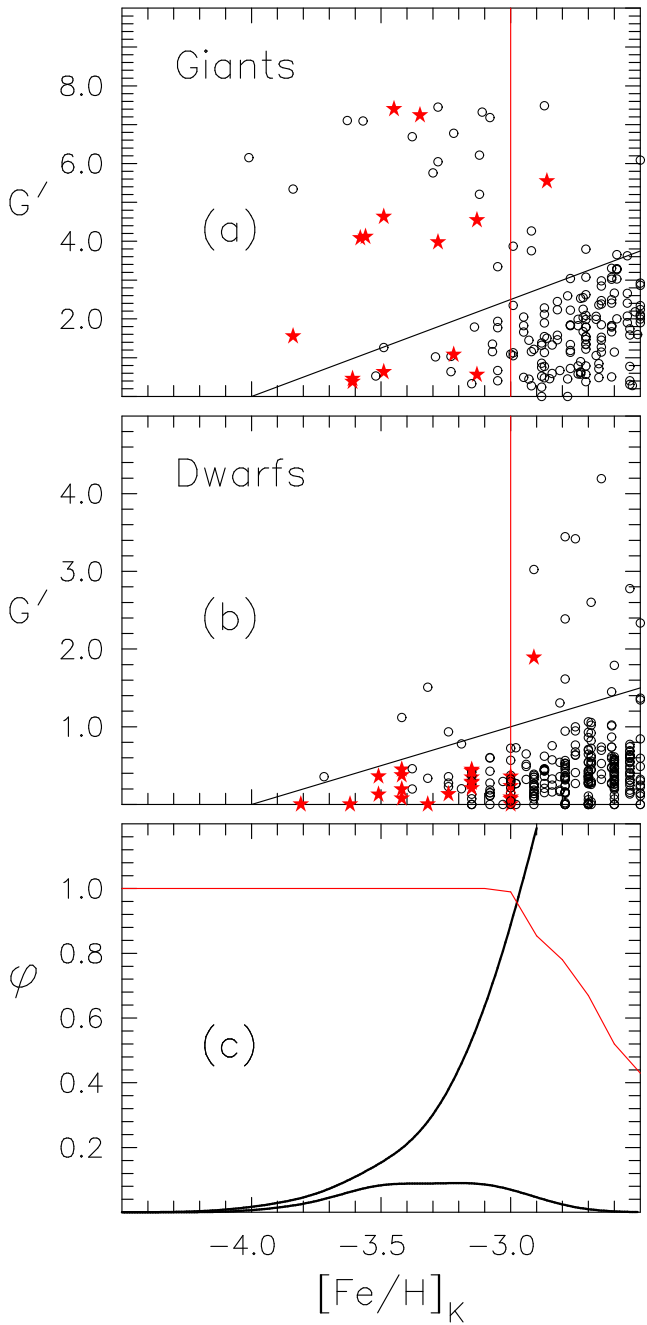}

  \caption{\label{Fig:2.3mdf}\small {\gp} vs. {\fehk} for (a) giants
    and (b) dwarfs in the Hamburg/ESO survey, measured on
    medium-resolution spectra obtained with the ANU 2.3\,m Telescope.
    Filled (red) symbols represent stars for which high-resolution,
    high-$S/N$ spectra have been obtained (in both this and other
    investigations).  Panel (c) presents the generalized histograms
    (Gaussian kernel with $\sigma$ = 0.15) of the MDF for stars in the
    medium-resolution (upper thick line) and high-resolution (lower
    thick line) samples. The scale of the ordinate is arbitrary, but
    is normalized so that the areas under the two distributions have
    the same proportionality to the numbers of stars in the samples.
    The principally horizontal, thin line represents the completeness
    function of the MDF obtained from medium-resolution spectroscopy
    of the HES, as discussed in the text. (The ordinate runs between
    0. and 1.15.)}

\end{center}
\end{figure}

In Figure~\ref{Fig:2.3mdf}, we present stars having {\fehk} $<$ --2.5
in the {\gp} (carbon-sensitive index) vs. {\fehk} plane, where the top
panel contains objects with {\bvz} $>$ 0.55 (principally giants) and
the middle panel shows those with {\bvz} $< $ 0.55 (principally
dwarfs).  The sloping lines in the two panels are arbitrarily chosen
to separate potentially carbon-normal and carbon-rich objects, while
the vertical lines are included to emphasize the region of immediate
interest for the present investigation -- {\fehk} $<$ --3.0.  More
metal-poor than this limit our 2.3m sample survey contains 109 stars.
In Figure~\ref{Fig:2.3mdf} we identify those objects with {\fehk}
$\la$ --3.0 for which high-resolution, high-$S/N$ data have been
obtained in the present work, together with those of \citet{norris07}
and \citet{garcia08}, as filled star symbols.  Open circles represent
stars for which high-resolution data are not yet available.

The bottom panel in Figure~\ref{Fig:2.3mdf} presents the MDF of all of
the stars in the upper two panels, where the upper thick line
represents the distribution for the complete sample of stars from the
2.3\,m HES sample described above, and the lower thick line shows that
of only the stars having high-resolution data.  The thin, mostly
horizontal line in the figure shows the completeness function from
\citet{schoerck09} and \citet{li10} for the detection of metal-poor
stars in medium-resolution investigations of the HES: below [Fe/H] =
--3.0, the HES is essentially complete.  Consideration of the thick
lines in the figure shows quite clearly the strong bias we have
introduced into the completeness function by our emphasis on observing
the most metal-poor stars.  We shall bear this in mind in subsequent
discussions of the MDF at the lowest metallicities.

\subsection{Supplementary Selection}

We augmented our program sample by the inclusion of eight stars from other
sources.  These comprise four stars from the Sloan Digital Sky Survey
(SDSS; \citealt{york00}), three dwarfs (BS~16545-089 and HE~1346--0427
from \citet{cohen04}, and HE~0945--1435 from \citet{garcia08}), and
the red giant CS~30336-049 (\citealp {tanner06, lai08}).

\subsection{The Sample}

Our total sample thus comprises 38 stars -- 30 from the 2.3m survey
described in Section~\ref{Sec:2.3m}, four from the SDSS, BS~16545-089
and CS~30336-049 for the HK survey, and HE~0945--1435 and
HE~1346--0427 from the HES.  The basis data for these stars are
presented in Table~\ref{Tab:sample}.  Columns (1)--(3) contain the
star names and coordinates, columns (4)--(6) contain B$_{\rm HES}$,
{\bvz}, and the source of the color, and columns (7)--(10) present
{\kp}, {\gp}, {\hp}, and {\fehk}, respectively.  (We note that the
first four entries in the table are based on spectra taken from the
SDSS archives.)  For completeness and comparison purposes, column (11)
presents values of [Fe/H] from Paper II, which result from the model
atmosphere analysis of the high-resolution, high-$S/N$ spectra
presented here.  Of these stars, 34 are original to the present work.

\begin{deluxetable*}{lccccrrrrrr}
\tabletypesize{\scriptsize}
\tablecolumns{11}
\tablewidth{0pt}
\tablecaption{\label{Tab:sample} THE SAMPLE OF 38 EXTREMELY METAL-POOR STARS}
\tablehead{
\colhead{Star} & {RA} & {Dec} & {$B_{\rm HES}$} & {\bvz} & {S\tablenotemark{a}} & {\kp} & {\gp} & {\hp} & {\fehk} & {[Fe/H]} \\ 
\colhead{} & {(2000)} & {(2000)} & {} & {} & {} & {({\AA})} & {({\AA})} & {({\AA})} & {} & {} \\ 
\colhead{(1)} & {(2)} & {(3)} & {(4)} & {(5)} & {(6)} & {(7)} & {(8)} & {(9)} & {(10)} & {(11)}  
  }
\startdata
52972-1213-507\tablenotemark{b} & 09 18 49.9 &   +37 44 27 &  16.2\tablenotemark{c}  &  0.42 &  1 &  0.81 &   3.88 &   3.77 &   --3.32 &  $-$2.98 \\
53327-2044-515\tablenotemark{b} & 01 40 36.2 &   +23 44 58 &  15.6\tablenotemark{c}  &  0.51 &  1 &  0.90 &   0.50 &   2.69 &   --3.49 &  $-$4.04\tablenotemark{d} \\
53436-1996-093\tablenotemark{b} & 11 28 13.5 &   +38 41 49 &  15.8\tablenotemark{c}  &  0.38 &  1 &  0.45 &   0.10 &   4.48 &   --3.81 &  $-$3.53 \\
54142-2667-094\tablenotemark{b} & 08 51 36.7 &   +10 18 03 &  15.4\tablenotemark{c}  &  0.38 &  1 &  0.92 &   0.00 &   4.39 &   --3.24 &  $-$2.96 \\
BS~16545-089    & 11 24 27.6 &   +36 50 28 &  14.8\tablenotemark{c}  &  0.38 &  2 &   ... &    ... &    ... &      ... &  $-$3.44 \\
CS~30336-049    & 20 45 23.5 & $-$28 42 36 &  14.9\tablenotemark{c}  &  0.86 &  3 &  1.68 &   0.21 &   0.80 &   --3.82 &  $-$4.10 \\
HE~0049--3948   & 00 52 13.4 & $-$39 32 37 &  15.6  &  0.35 &  1 &  0.50 &   0.08 &   5.11 &  $-$3.42 &  $-$3.68 \\
HE~0057--5959   & 00 59 54.1 & $-$59 43 30 &  16.4  &  0.61 &  1 &  0.85 &   0.38 &   1.68 &  $-$3.61 &  $-$4.08 \\
HE~0102--1213   & 01 05 28.0 & $-$11 57 29 &  14.1  &  0.43 &  1 &  1.03 &   0.29 &   3.66 &  $-$3.15 &  $-$3.28 \\
HE~0146--1548   & 01 48 34.7 & $-$15 33 25 &  16.5  &  0.89 &  4 &  3.09 &   4.08 &   0.78 &  $-$3.58 &  $-$3.46 \\
HE~0207--1423   & 02 10 00.7 & $-$14 09 12 &  15.8  &  0.82 &  4 &  2.63 &   7.40 &   2.29 &  $-$3.45 &  $-$2.95 \\
HE~0228--4047   & 02 30 33.6 & $-$40 33 56 &  15.5  &  0.34 &  1 &  0.41 &   0.00 &   5.34 &  $-$3.62 &  $-$3.75\tablenotemark{d} \\
HE~0231--6025   & 02 32 30.7 & $-$60 12 11 &  15.1  &  0.35 &  1 &  0.71 &   0.45 &   5.03 &  $-$3.42 &  $-$3.10 \\
HE~0253--1331   & 02 56 06.6 & $-$13 19 27 &  15.9  &  0.36 &  1 &  0.89 &   0.43 &   4.96 &  $-$3.15 &  $-$3.01 \\
HE~0314--1739   & 03 17 01.8 & $-$17 28 56 &  16.5  &  0.36 &  1 &  0.64 &   0.38 &   5.00 &  $-$3.42 &  $-$2.86 \\
HE~0355--3728   & 03 57 23.0 & $-$37 20 23 &  16.3  &  0.41 &  1 &  0.90 &   0.00 &   3.94 &  $-$3.32 &  $-$3.41\tablenotemark{d} \\
HE~0945--1435   & 09 47 50.7 & $-$14 49 07 &  15.1  &  0.39 &  1 &  0.69 &   0.13 &   4.27 &  $-$3.51 &  $-$3.78\tablenotemark{d} \\
HE~1055+0104    & 10 58 04.4 &   +00 48 36 &  14.6  &  0.40 &  1 &  1.23 &   0.36 &   4.21 &  $-$3.00 &  $-$2.88\tablenotemark{d} \\
HE~1116--0054   & 11 18 47.8 & $-$01 11 20 &  16.5  &  0.39 &  1 &  0.45 &   0.00 &   4.28 &  $-$3.81 &  $-$3.48\tablenotemark{d} \\
HE~1142--1422   & 11 44 59.3 & $-$14 38 49 &  14.7  &  0.40 &  1 &  1.17 &   0.00 &   4.12 &  $-$3.00 &  $-$2.84 \\
HE~1201--1512   & 12 03 37.1 & $-$15 29 32 &  14.2  &  0.50 &  1 &  0.93 &   0.63 &   2.73 &  $-$3.49 &  $-$3.89\tablenotemark{d} \\
HE~1204--0744   & 12 06 46.2 & $-$08 00 43 &  15.2  &  0.37 &  1 &  0.92 &   0.34 &   4.60 &  $-$3.15 &  $-$2.71 \\
HE~1207--3108   & 12 09 54.1 & $-$31 25 11 &  12.9  &  0.63 &  1 &  1.69 &   0.56 &   1.53 &  $-$3.13 &  $-$2.70 \\
HE~1320--2952   & 13 22 55.0 & $-$30 08 06 &  14.5  &  0.68 &  1 &  1.82 &   1.08 &   1.14 &  $-$3.22 &  $-$3.69 \\
HE~1346--0427   & 13 49 25.2 & $-$04 42 15 &  14.5  &  0.45 &  4 &   ... &    ... &    ... &      ... &  $-$3.58\tablenotemark{d} \\
HE~1402--0523   & 14 04 37.9 & $-$05 38 13 &  15.8  &  0.38 &  1 &  0.86 &   0.13 &   4.50 &  $-$3.24 &  $-$3.18\tablenotemark{d} \\
HE~1506--0113   & 15 09 14.3 & $-$01 24 57 &  15.4  &  0.64 &  1 &  1.48 &   3.97 &   1.42 &  $-$3.28 &  $-$3.54 \\
HE~2020--5228   & 20 24 17.0 & $-$52 19 03 &  16.3  &  0.38 &  1 &  1.00 &   0.07 &   4.48 &  $-$3.00 &  $-$2.93 \\
HE~2032--5633   & 20 36 24.9 & $-$56 23 05 &  15.2  &  0.36 &  1 &  0.75 &   0.21 &   4.97 &  $-$3.15 &  $-$3.63 \\
HE~2047--5612   & 20 51 22.2 & $-$56 00 52 &  15.5  &  0.44 &  1 &  1.09 &   0.28 &   3.54 &  $-$3.15 &  $-$3.14 \\
HE~2135--1924   & 21 38 04.7 & $-$19 11 04 &  15.7  &  0.37 &  1 &  0.60 &   0.19 &   4.69 &  $-$3.42 &  $-$3.31 \\
HE~2136--6030   & 21 40 39.7 & $-$60 16 27 &  15.9  &  0.38 &  1 &  1.21 &   0.23 &   4.54 &  $-$3.00 &  $-$2.88 \\
HE~2139--5432   & 21 42 42.5 & $-$54 18 43 &  15.9  &  0.51 &  4 &  0.83 &   4.63 &   3.49 &  $-$3.49 &  $-$4.02 \\
HE~2141--0726   & 21 44 06.6 & $-$07 12 49 &  15.7  &  0.38 &  1 &  1.21 &   0.09 &   4.54 &  $-$3.00 &  $-$2.72 \\
HE~2142--5656   & 21 46 20.5 & $-$56 42 18 &  14.4  &  0.71 &  4 &  4.06 &   5.54 &   1.08 &  $-$2.86 &  $-$2.87 \\
HE~2202--4831   & 22 06 06.0 & $-$48 16 53 &  16.3  &  0.71 &  4 &  1.88 &   7.24 &   2.33 &  $-$3.35 &  $-$2.78 \\
HE~2246--2410   & 22 48 59.7 & $-$23 54 38 &  15.7  &  0.35 &  1 &  0.92 &   0.44 &   5.12 &  $-$3.15 &  $-$2.96 \\
HE~2247--7400   & 22 51 19.6 & $-$73 44 24 &  14.4  &  1.02 &  4 &  4.49 &   4.54 &   1.00 &  $-$3.13 &  $-$2.87 \\
\enddata
\tablenotetext{a}{Source of {\bvz}: 1 = {\hp}, see Section~\ref{Sec:2.3m}; 2 = \citet{cohen04}; 3 = \citet{norris99}; 4 = {\bvhes}, see Section~\ref{Sec:2.3m}}
\tablenotetext{b}{SDSS  ``MJD-plug plate-fiber'' nomenclature}
\tablenotetext{c}{$B$ from SDSS (rows (1)--(4)); \citet{cohen08} (row (5), adopting color from this table); and \citet{norris99} (row (6))} 
\tablenotetext{d}{Average of dwarf and subgiant values of [Fe/H] from Table 1 of Paper II}
\end{deluxetable*}

\section{HIGH RESOLUTION SPECTROSCOPY}

\subsection{Magellan, Keck, and VLT Spectra} \label{Sec:spectra}

High-resolution, high-$S/N$ spectra of the 38 program stars in
Table~\ref{Tab:sample} were obtained with the Magellan/MIKE, the
Keck/HIRES, and VLT/UVES telescope/spectrograph
combinations during 2007--2008.  Details of the observing sessions and
instrumental set-ups are presented in Table~\ref{Tab:obslog}, where
columns (1)--(4) present the telescope/spectrograph combinations,
observing dates, wavelength ranges, and resolving powers
(R = $\lambda/(\Delta/\lambda)$), respectively.  

\begin{deluxetable*}{llcc}
\tabletypesize{\small}
\tablecolumns{4}
\tablewidth{0pt}
\tablecaption{\label{Tab:obslog} LOG OF HIGH-RESOLUTION SPECTROSCOPIC OBSERVATIONS}
\tablehead{
\colhead{Telescope/} & 
\colhead{Date}    & 
\colhead{Wavelength range} & 
\colhead{Resolving power} \\
\colhead{Spectrograph}    & {}    & {(\AA)} & {}   \\
\colhead{(1)}             & {(2)} & {(3)}   & {(4)}
  }
\startdata
Magellan/MIKE & 2007 Jun 20--23             &  3300--4900  & 39000 \\ 
              &                             &  4900--9400  & 31000 \\
              & 2007 Dec 21--22             &  3300--4900  & 37000 \\
              &                             &  4900--9400  & 30000 \\ 
              & 2008 Sep 6--7               &  3300--4900  & 36000 \\
              &                             &  4900--9400  & 30000 \\
              &                             &              &       \\
Keck/HIRES    & 2007 Nov 30 -- Dec 1        &  3720--4600  & 48000 \\
              &                             &  4660--5600  & 48000 \\
              &                             &  5660--6540  & 48000 \\
              & 2008 Mar 25--28             &  4020--4660  & 49000 \\
              &                             &  4750--6190  & 49000 \\
              &                             &  6340--7760  & 49000 \\
              &                             &              &       \\
VLT/UVES      & 2008 Jul 3--24              &  3300--4520  & 40000 \\
              & (2.7 hrs Service Observing) &  4790--5750  & 40000 \\
              &                             &  5840--6800  & 40000 \\
\enddata
\end{deluxetable*}

The Magellan and Keck observations were obtained in Visitor Mode,
during which data were also obtained of ``standard'' metal-poor stars
for comparison purposes, together with quartz-iodine and ThAr lamps
for flat-fielding and wavelength calibration.  We tailored data
acquisition for the program stars to obtain $S/N \sim 100$ per pixel at
4500\,{\AA} for the most metal-poor stars in our sample. To do this we
began our observations of each star in ``snapshot'' mode with a
minimal number ($\sim$1--3) of 1800~sec exposures, which we reduced in
real time to permit us, by comparison with our library of
high-resolution extremely metal-poor stars, to give priority to the
more metal-poor objects.

We then obtained additional exposures as necessary to yield
higher $S/N$ for the most interesting objects.  The reader will see
this reflected to some extent in the $S/N$ values reported below.
The data were processed with standard IRAF{\footnote{IRAF is
    distributed by the National Optical Astronomy Observatories, which
    are operated by the Association of Universities for Research in
    Astronomy, Inc., under cooperative agreement with the National
    Science Foundation.} procedures (supplemented by the cosmic-ray
  removal algorithm of \citealp{pych04}) to obtain flat-fielded,
  wavelength-calibrated, comic-ray-corrected, co-added, and
  continuum-normalized spectra (see, e.g., \citealp{yong03}).

Four exposures were obtained for HE~1506--0113 in Service Mode with
the VLT/UVES system, each having an integration time of 2500 sec, with
spectrograph settings as used in our investigation of HE~0557--4840
\citep{norris07}.  These four ESO pipeline-reduced spectra were
co-added to produce the final spectrum, which was then continuum
normalized.

Examples of the high-resolution spectra are presented in the lower
panels of Figure~\ref{Fig:spectra}, which cover the wavelength range
3900--4000\,{\AA}, together in the upper panels with the corresponding
medium-resolution spectra (3800--4600\,{\AA}) described in
Section~\ref{Sec:2.3m}.  The $S/N$ (per ~0.017\,{\AA} pixel at
4500\,{\AA}) of the final high-resolution spectra are presented in the
first row of each of Tables~\ref{Tab:table3}--\ref{Tab:table6}.  The
range in $S/N$ for the total sample is 23--210, with median 61, while
for stars with high-resolution abundances [Fe/H] $<$ --3.5 the median
$S/N$ is 118.  For the latter group, we note for the record that the
mean $B$ magnitude is 15.3.

\begin{center}                                                                                                          
\begin{deluxetable*}{cccrrrrrrrrrr}                                                                                      
\tabletypesize{\tiny}                                                                                                   
\tablecolumns{13}                                                                                                       
\tablewidth{0pt}                                                                                                        
\tablecaption{\label{Tab:table3} ATOMIC DATA AND EQUIVALENT WIDTHS (m{\AA}) FOR PROGRAM STARS }                       
\tablehead{                                                                                                             
\colhead{Wavelength} & {Species} & {$\chi$} & {log$gf$} & {52972}      & {53327}      & {53436}      & {54142}      & {BS~16545} & {CS~30336} & {HE~0049} & {HE~0057} & {HE~0102} \\
{(\AA)}   & &{($\rm eV$)} &                             & {-1213-507}  & {-2044-515}  & {-1996-093}  & -2667{-094}  & {-089}     & {-049}     & {--3948}  & {--5959}  & {--1213} \\
{(1)}     & {(2)} & {(3)} & {(4)}                       & {(5)}        & {(6)}        & {(7)}        & {(8)}        & {(9)}      & {(10)}     & {(11)}    & {(12)}    & {(13)}
}                                                                                                                       
\startdata                                                                                                              
$S/N$\tablenotemark{a}  &  &  &  &    28 &     51 &    129 &     47 &    180 &    126 &    149 &     91 &     81 \\
W(min)\,(m{\AA})  &  &  &  &    12 &      7 &      6 &      8 &      5 &      6 &      6 &      6 &      5 \\\\
 5889.95 &   11.0 &   0.00 &    0.11 &  183.0 &    ...  &    ...  &    ...  &   20.2  &   70.4  &   10.6  &  156.0  &    ...  \\
 5895.92 &   11.0 &   0.00 &   $-$0.19 &  105.0  &   17.8  &   10.9  &    ...  &   11.6  &   50.6  &   12.6  &  129.0  &   32.7  \\
 3829.36 &   12.0 &   2.71 &   $-$0.21 &   93.0  &   65.4  &    ...  &   88.7  &    ...  &   93.6  &   43.8  &   86.1  &   78.5  \\
 3832.30 &   12.0 &   2.71 &    0.15 &    ...  &    ...  &    ...  &    ...  &    ...  &  106.5  &    ...  &  106.5  &    ...  \\
 3838.29 &   12.0 &   2.72 &    0.41 &    ...  &    ...  &    ...  &    ...  &    ...  &  117.5  &    ...  &  109.5  &    ...  
\enddata                                                                                                                
\tablenotetext{a}{$S/N$ per $\sim$0.17\,{\AA}~pixel at 4500\,{\AA}}                                                      
\tablenotetext{b}{These lines produce discrepant abundances and are not included in the results reported in Paper II}   
\tablerefs{
Note. Table 3 is published in its entirety in the electronic edition of The Astrophysical Journal. A portion is shown here for guidance regarding its form and content.
}
\end{deluxetable*}                                                                                                       

\end{center}

\begin{center}                                                                                                          
\begin{deluxetable*}{cccrrrrrrrrrrr}                                                                                     
\tabletypesize{\tiny}                                                                                                   
\tablecolumns{14}                                                                                                       
\tablewidth{0pt}                                                                                                        
\tablecaption{\label{Tab:table4} ATOMIC DATA AND EQUIVALENT WIDTHS (m{\AA}) FOR PROGRAM STARS}                       
\tablehead{                                                                                                             
\colhead{Wavelength} & {Species} & {$\chi$} & {log$gf$} & {HE~0146} & {HE~0207} & {HE~0228} & {HE~0231} & {HE~0253} & {HE~0314} & {HE~0355} & {HE~0945} & {HE~1055} & {HE~1116} \\
{(\AA)}   & &{($\rm eV$)} &                              & {--1548} & {--1423}  & {--4047}  & {--6025}  & {--1331}  & {--1739}  & {--3728}  & {--1435}  & {+0104 }  & {--0054} \\
{(1)} & {(2)} & {(3)} & {(4)}                            & {(5)}    & {(6)}     & {(7)}     & {(8)}     & {(9)}     & {(10)}    & {(11)}    & {(12)}    & {(13)}    & {(14)}
}                                                                                                                       
\startdata                                                                                                              
$S/N$\tablenotemark{a}  &  &  &  &    34 &     23 &    118 &     42 &     41 &     43 &     29 &    158 &     49 &    128 \\
W(min)\,(m{\AA})  &  &  &  &    14 &     13 &      6 &     13 &     10 &      9 &     20 &      5 &      9 &      7 \\\\
 5889.95 &   11.0 &   0.00 &    0.11 &  186.0  &  138.5  &    ...  &   44.2  &   45.1  &   29.9  &    ...  &    ...  &   67.1  &   16.4  \\
 5895.92 &   11.0 &   0.00 &   $-$0.19 &  154.5  &    ...  &    ...  &   37.7  &   23.4  &    ...  &    ...  &    ...  &   39.6  &   10.6  \\
 3829.36 &   12.0 &   2.71 &   $-$0.21 &    ...  &    ...  &   43.2  &  100.2  &   88.8  &   78.4  &    ...  &    ...  &   98.1  &   55.0  \\
 3832.30 &   12.0 &   2.71 &    0.15 &    ...  &    ...  &    ...  &    ...  &    ...  &    ...  &    ...  &    ...  &    ...  &    ...  \\
 3838.29 &   12.0 &   2.72 &    0.41 &    ...  &    ...  &    ...  &    ...  &    ...  &    ...  &    ...  &    ...  &    ...  &    ...  \\
\enddata                                                                                                                
\tablenotetext{a}{$S/N$ per $\sim$0.17\,{\AA}~pixel at 4500\,{\AA}}                                                      
\tablenotetext{b}{These lines produce discrepant abundances and are not included in the results reported in Paper II}   
\tablerefs{
Note. Table 4 is published in its entirety in the electronic edition of The Astrophysical Journal. A portion is shown here for guidance regarding its form and content.
}
\end{deluxetable*}                                                                                                       
\end{center}

\begin{center}                                                                                                          
\begin{deluxetable*}{cccrrrrrrrrrrr}                                                                                     
\tabletypesize{\tiny}                                                                                                   
\tablecolumns{14}                                                                                                       
\tablewidth{0pt}                                                                                                        
\tablecaption{\label{Tab:table5} ATOMIC DATA AND EQUIVALENT WIDTHS (m{\AA}) FOR PROGRAM STARS}                        
\tablehead{                                                                                                             
\colhead{Wavelength} & {Species} & {$\chi$} & {log$gf$} & {HE~1142} & {HE~1201} & {HE~1204} & {HE~1207} & {HE~1320} & {HE~1346} & {HE~1402} & {HE~1506} & {HE~2020} & {HE~2032} \\
{(\AA)}   & &{($\rm eV$)} &                             & {--1422}  & {--1512}  & {--0744}  & {--3108}  & {--2952}  & {--0427}  & {--0523}  & {--0113}  & {--5228}  & {--5633} \\
{(1)} & {(2)} & {(3)} & {(4)}                           & {(5)}     & {(6)}     & {(7)}     & {(8)}     & {(9)}     & {(10)}    & {(11)}    & {(12)}    &  {(13)}   & {(14)}
}                                                                                                                       
\startdata                                                                                                              
$S/N$\tablenotemark{a}  &  &  &  &    62 &    219 &     27 &     66 &     61 &    206 &     77 &     64 &     37 &     56 \\
W(min)\,(m{\AA})  &  &  &  &    15 &      5 &     16 &     10 &     10 &      5 &      8 &      5 &     13 &     10 \\\\
 5889.95 &   11.0 &   0.00 &    0.11 &   87.9  &   17.7  &   67.8  &  156.0  &   79.5  &   19.8  &   29.7  &  185.0  &    ...  &   23.5  \\
 5895.92 &   11.0 &   0.00 &   $-$0.19 &   53.7  &    8.5  &    ...  &  105.0  &   55.8  &   13.7  &   25.9  &  152.0  &    ...  &    9.9  \\
 3829.36 &   12.0 &   2.71 &   $-$0.21 &  139.5  &   60.7  &    ...  &  101.8  &  100.7  &    ...  &   86.7  &  138.0  &   85.5  &   59.6  \\
 3832.30 &   12.0 &   2.71 &    0.15 &    ...  &    ...  &    ...  &  128.5  &  123.0  &    ...  &    ...  &    ...  &  116.5  &    ...  \\
 3838.29 &   12.0 &   2.72 &    0.41 &    ...  &    ...  &    ...  &  143.0  &  136.5  &    ...  &    ...  &    ...  &  121.5  &    ...  \\
\enddata                                                                                                                
\tablenotetext{a}{$S/N$ per $\sim$0.17\,{\AA}~pixel at 4500\,{\AA}}                                                      
\tablenotetext{b}{These lines produce discrepant abundances and are not included in the results reported in Paper II}   
\tablerefs{
Note. Table 5 is published in its entirety in the electronic edition of The Astrophysical Journal. A portion is shown here for guidance regarding its form and content.
}
\end{deluxetable*}                                                                               

\end{center}

\begin{center}                                                                                                          
\begin{deluxetable*}{cccrrrrrrrrrr}                                                                                      
\tabletypesize{\tiny}                                                                                                   
\tablecolumns{13}                                                                                                       
\tablewidth{0pt}                                                                                                        
\tablecaption{\label{Tab:table6} ATOMIC DATA AND EQUIVALENT WIDTHS (m{\AA}) FOR PROGRAM STARS}                       
\tablehead{                                                                                                             
\colhead{Wavelength} & {Species} & {$\chi$} & {log$gf$} & {HE~2047} & {HE~2135} & {HE~2136} & {HE~2139} & {HE~2141} & {HE~2142} & {HE~2202} & {HE~2246} & {HE~2247} \\
{(\AA)}   & &{($\rm eV$)} &                             & {--5612}  & {--1924}  & {--6030}  & {--5432}  & {--0726}  & {--5656}  & {--4831}  & {--2410}  & {--7400} \\
{(1)} & {(2)} & {(3)} & {(4)}                           & {(5)}     & {(6)}     & {(7)}     & {(8)}     & {(9)}     & {(10)}    & {(11)}    & {(12)}    & {(13)}
}                                                                                                                       
\startdata                                                                                                              
$S/N$\tablenotemark{a}  &  &  &  &    63 &     75 &     31 &     86 &     44 &     48 &     37 &     61 &     23 \\
W(min)\,(m{\AA})  &  &  &  &    10 &      7 &     15 &      8 &     10 &     15 &     15 &      8 &     20 \\\\
 5889.95 &   11.0 &   0.00 &    0.11 &   44.2  &   17.6 &   38.7  &  146.0  &   45.2  &  187.0  &    ...  &  142.0  &    ...  \\
 5895.92 &   11.0 &   0.00 &   $-$0.19 &   22.2  &   36.2  &   24.9  &  130.0  &   26.5  &  166.5  &  192.5  &  120.0  &  170.5  \\
 3829.36 &   12.0 &   2.71 &   $-$0.21 &   68.7  &   57.5  &   70.6  &  136.0  &   88.3  &  166.5  &    ...  &   85.3  &    ...  \\
 3832.30 &   12.0 &   2.71 &    0.15 &    ...  &    ...  &  120.0  &    ...  &    ...  &    ...  &    ...  &    ...  &    ...  \\
 3838.29 &   12.0 &   2.72 &    0.41 &    ...  &    ...  &  119.0  &    ...  &    ...  &    ...  &    ...  &    ...  &    ...  \\
\enddata                                                                                                                
\tablenotetext{a}{$S/N$ per $\sim$0.17\,{\AA}~pixel at 4500\,{\AA}}                                                      
\tablenotetext{b}{These lines produce discrepant abundances and are not included in the results reported in Paper II}   
\tablerefs{
Note. Table 6 is published in its entirety in the electronic edition of The Astrophysical Journal. A portion is shown here for guidance regarding its form and content.
}
\end{deluxetable*}                                                                                                       
\end{center}

\begin{figure}[!tbp]
\figurenum{2}
\begin{center}
\includegraphics[width=8.5cm,angle=0]{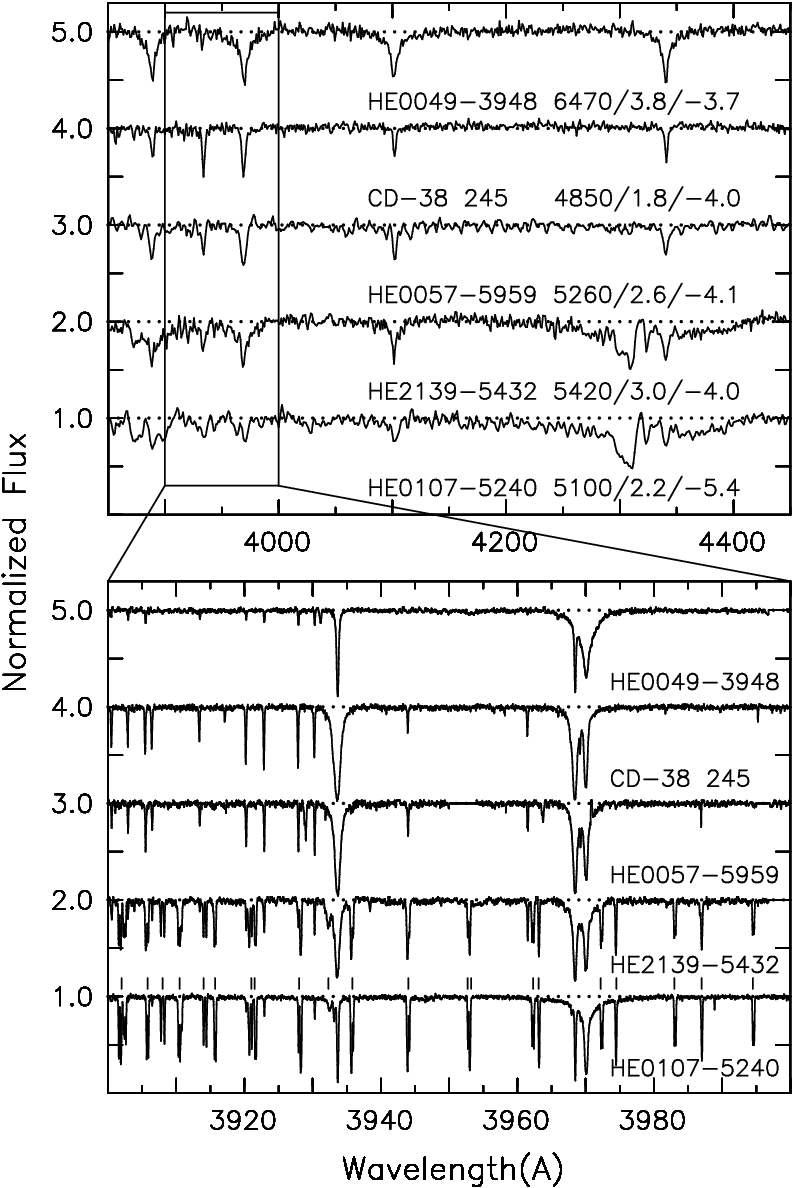}

  \caption{\label{Fig:spectra}\small Spectra of some of the most
    metal-poor stars. HE~0049--3948, HE~0057--5959, and HE~2139--5432
    are from the present work, while {\cd} and {\hen} derive from
    \citet{bessell84} and \citet{christlieb04}, respectively, and are
    shown for comparison purposes.  The upper panel presents
    medium-resolution (R $\sim$ 1600) spectra obtained with the 2.3\,m
    Telescope and cover the range 3850--4450\,{\AA}.  The numbers in
    the panel represent {\teff}/{\logg}/[Fe/H] from analyses of
    high-resolution, high-$S/N$ spectra by \citet{christlieb04},
    \citet{norris01}, the present work, and Paper II.  The lower panel
    shows spectra of the same stars at R $\sim$ 40000 on the range
    3900--4000\,{\AA}.  The reader will note that while the
    Ca\,II\,H\,\&\,K lines (at 3968.4 and 3933.6\,{\AA}, respectively)
    are very weak in the most metal-poor giant, {\hen}, many more
    lines have appeared in its spectrum.  These are features of CH
    (the positions of which are indicated immediately above the
    spectrum) resulting from an extremely large overabundance of
    carbon relative to iron in this object.  We note that
    HE~2139--5432 is similar to {\hen} in this respect, but not as
    metal-poor.}

\end{center}
\end{figure}

\subsection{Equivalent Widths}

Equivalent widths have been measured from these spectra as described
in \citet{norris10c,norris10a}, to whom we refer the reader for
details.  As noted there, we begin with the line list of
\citet{cayrel04}, which we supplement here with lines of Sr~II and
Ba~II\footnote{We note for completeness that we also sought,
  unsuccessfully, to measure lines of La II, Ce II, Nd II, and Eu
  II.}.  We also recall that the spectra of carbon-rich stars can be
heavily contaminated by CH lines.  To minimize this effect we followed
the spectrum synthesis technique of \citet{norris10a} to determine
when the blending of CH lines with atomic features would contaminate
the latter.  For stars having {\gp} $>$ 3.5\,{\AA}, we did not measure
atomic features susceptible to such contamination.  We also excluded
lines in regions with observed C$_{2}$ absorption in these objects.

For 37 of the program objects, each spectrum was measured
independently by J.~E.~N. and D.~Y., using techniques described by
\citet{norris01} and \citet{yong08}\footnote{The line strengths of
  HE~1506-0113 were measured by only J.~E.~N.}.  In general, the
agreement between the results was excellent, an example of which is
shown for a representative program star in
Figure~\ref{Fig:equivalent_widths}(a).  For the 37 objects, the median
RMS scatter between the two authors was 2.7 m{\AA}.  We also estimated
the smallest equivalent width that we could reliably measure in each
of the spectra, which we present in the second row of each of
Tables~\ref{Tab:table3}--\ref{Tab:table6}.  (The median smallest
equivalent width for our sample is 10\,m{\AA}, while for stars having
[Fe/H] $<$ --3.5 it is 6\,m{\AA} -- driven by our emphasis on
obtaining higher $S/N$ for the more metal-poor stars in our sample.)
Our adopted equivalent widths are the simple average of the results of
J.E.N. and D.Y. for each star.  These are presented in
Tables~\ref{Tab:table3}--\ref{Tab:table6}, which contain line
strengths for 14 elements in the wavelength range
$\sim$3750--6500\,{\AA}.  Columns (1)--(4) contain line
identification, lower excitation potential ($\chi$), and {\loggf}
value, respectively, for 191 unblended lines suitable for model
atmosphere abundance analysis, taken from \citet[Table~3]{cayrel04},
together with values from the literature for Sr~II and
Ba~II\footnote{The {\loggf} values are from \citet{pinnington95},
  \citet{gallagher67}, and VALD (see \citealt{kupka99}).}.  The
remainder of the table is populated by equivalent widths.  Some of the
lines lead to significantly discrepant abundances compared with those
obtained from other lines of the same species.  We flag these values
in Tables~\ref{Tab:table3}--\ref{Tab:table6}, and have excluded them
from the analysis in Paper II.

\begin{figure}[!tbp]
\figurenum{3}
\begin{center}
\includegraphics[width=6.0cm,angle=0]{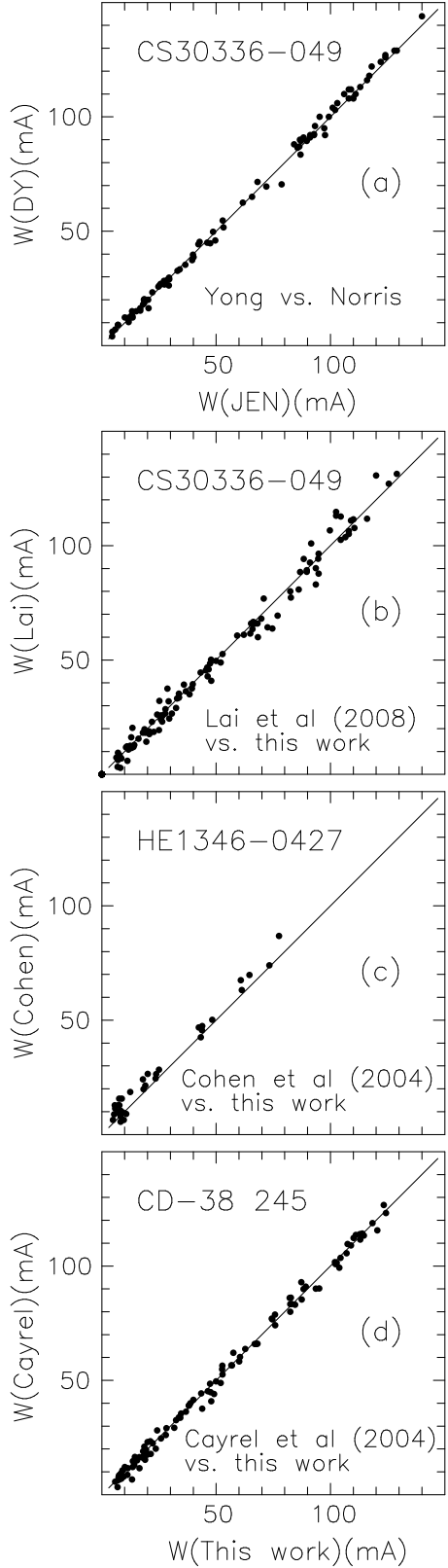}

  \caption{\label{Fig:equivalent_widths}\small Comparison of
    equivalent widths (a) measured in the present work by J.~E.~N. and
    D.~Y., and (b)--(d) between the present work and that of
    \citep{lai08}, \citep{cohen04}, and \citep{cayrel04}.}

\end{center}
\end{figure}

We have compared our results with those of other workers.
Figure~\ref{Fig:equivalent_widths}(b) and (c) compare the present
equivalent widths with those of \citet{lai08} (CS~30336-049) and
\citet{cohen04} (HE~1346--0427), for which the RMS scatters are 3.0
and 4.4 m{\AA}, respectively. We have also obtained data for
well-observed stars in the present program and found good agreement
with the results of others.  As an example,
Figure~\ref{Fig:equivalent_widths}(d) compares our equivalent widths
with those of \citet{cayrel04} for {\cd}, where the RMS scatter is
2.4~m{\AA}.

\subsection{Radial Velocities}

Radial velocities were measured from the high-resolution spectra (for
all but one program star, HE~1506--0113) using the FXCOR task in
IRAF. We cross-correlated individual spectra with a template spectrum
obtained with the same set-up on the same observing run (generally a
high-$S/N$ exposure of a metal-poor standard)\footnote{Our templates
  were CS~22892-052 : 13.1~{\kms} (2007 June), BD$ -18^{\circ}\,
  5550$: --126.2~{\kms} (2007 November), BD$+9^{\circ}\,2190$:
  266.1~{\kms} (2007 December), BD$+9^{\circ}\,2190$: 266.1~{\kms}
  (2008 March), and HD140283: --170.4~{\kms} (2008 September).}. Given
the weakness of lines in these extremely metal-poor stars, only a
limited number of orders provided useful information. In these orders,
the peak of the cross-correlation function was fit with a Gaussian.
The velocities from multiple orders were averaged to obtain our final
radial velocities for each individual spectrum with a typical
uncertainty of 0.5~{\kms}, but with values as high as 3~{\kms}.  For
HE~1506--0113 the velocity was determined from the four individual VLT
spectra discussed in Section~\ref{Sec:spectra}, by cross-correlating
them with a model atmosphere synthetic spectrum as described by
\citet[Section 2.4]{norris10a}.

Table~\ref{Tab:average_velocities} presents average heliocentric
velocities obtained on each of our observing runs, together in
Table~\ref{Tab:individual_velocities} with velocities and the epoch of
observation for each individual observation.

\begin{deluxetable}{lcrc}
\tabletypesize{\tiny}
\tablecolumns{4} 
\tablewidth{0pc} 
\tablecaption{\label{Tab:average_velocities} RADIAL VELOCITIES \label{tab:rvs}} 
\tablehead{ 
\colhead{Star} & 
\colhead{Telescope\tablenotemark{a}/Date} & 
\colhead{V$_{\rm r}$\tablenotemark{b}} & 
\colhead{s.e.\tablenotemark{b}}\\ 
\colhead {} & 
\colhead {} & 
\colhead {(km s$^{-1}$)} & 
\colhead {(km s$^{-1}$)}\\ 
\colhead{(1)} & \colhead{(2)} & \colhead{(3)} & \colhead{(4)} 
}

\startdata 
52972-1213-507 & K Nov07 & $-$177.0 & 0.5 \\
53327-2044-515 & K Nov07 & $-$193.5 & 0.5 \\
53436-1996-093 & K Nov07 & $-$15.3 & 1.7 \\
               & K Mar08 & $-$17.2 & 0.5 \\
54142-2667-094 & K Nov07 & 43.4 & 0.7 \\
BS~16545-089 & K Mar08 & $-$161.1 & 0.5 \\
CS~30336-049 & M Jun07 & $-$236.6 & 0.8 \\
HE~0049--3948 & M Dec07 & 190.8 & 0.5 \\
            & M Sep08 & 190.6 & 0.9 \\
HE~0057--5959 & M Jun07 & 375.3 & 0.5 \\
HE~0102--1213 & K Nov07 & 90.5 & 0.5 \\
HE~0146--1548 & K Nov07 & $-$114.9 & 0.5 \\
HE~0207--1423 & K Nov07 & $-$209.6 & 0.5 \\
HE~0228--4047 & M Dec07 & 123.8 & 0.6 \\
            & M Sep08 & 128.5 & 1.5 \\
HE~0231--6025 & M Dec07 & 296.6 & 0.5 \\
HE~0253--1331 & K Nov07 & 25.2 & 0.5 \\
HE~0314--1739 & K Nov07 & 40.9 & 0.5 \\
HE~0355--3728 & M Dec07 & 82.0 & 1.8 \\
HE~0945--1435 & K Mar08 & 121.8 & 0.5 \\
HE~1055+0104 & K Nov07 & 324.6 & 0.5 \\
HE~1116--0054 & M Jun07 & 149.1 & 0.7 \\
            & K Nov07 & 148.3 & 1.3 \\
            & M Dec07 & 149.3 & 0.7 \\
            & K Mar08 & 148.3 & 0.8 \\
HE~1142--1422 & M Jun07 & $-$102.3 & 0.5 \\
HE~1201--1512 & M Jun07 & 238.8 & 0.5 \\
            & K Mar08 & 237.2 & 0.5 \\
HE~1204--0744 & M Jun07 & $-$49.7 & 2.7 \\
HE~1207--3108 & M Jun07 & $-$3.7 & 0.5 \\
HE~1320--2952 & M Jun07 & 390.0 & 0.5 \\
            & K Mar08 & 390.0 & 0.9 \\
HE~1346--0427 & K Mar08 & $-$47.8 & 0.5 \\
HE~1402--0523 & M Jun07 & $-$68.8 & 0.5 \\
HE~1506--0113 & V Jul08 & $-$137.1 & 0.3 \\
HE~2020--5228 & M Jun07 & $-$41.5 & 2.2 \\
HE~2032--5633 & M Jun07 & 260.5 & 0.5 \\
HE~2047--5612 & M Jun07 & $-$50.0 & 1.6 \\
HE~2135--1924 & K Nov07 & $-$252.9 & 1.6 \\
HE~2136--6030 & M Jun07 & 156.5 & 2.8 \\
HE~2139--5432 & M Dec07 & 113.4 & 0.7 \\
              & M Sep08 & 115.5 & 0.5 \\
HE~2141--0726 & K Nov07 & $-$38.1 & 0.5 \\
HE~2142--5656 & M Jun07 & 103.4 & 0.5 \\
HE~2202--4831 & M Jun07 & 56.2 & 0.5 \\
HE~2246--2410 & M Jun07 & $-$3.1 & 1.8 \\
HE~2247--7400 & M Jun07 & 5.7 & 0.5 \\
\enddata 
\tablenotetext{a}{K = Keck, M = Magellan, V = VLT}
\tablenotetext{b}{Heliocentric radial velocity and standard error of the mean}
\end{deluxetable}

\begin{deluxetable}{lccr}
\tabletypesize{\tiny}
\tablecolumns{4} 
\tablewidth{0pc} 
\tablecaption{\label{Tab:individual_velocities} INDIVIDUAL RADIAL VELOCITIES \label{tab:rvs2}} 
\tablehead{ 
\colhead{Star} & 
\colhead{Telescope\tablenotemark{a}/Date} & 
\colhead{Julian Date} & 
\colhead{V$_{\rm r}$\tablenotemark{b}} \\ 
\colhead{} & 
\colhead{} & 
\colhead{} & 
\colhead{(km s$^{-1}$)}\\  
\colhead{(1)} & \colhead{(2)} & \colhead{(3)} & \colhead{(4)} 
}
\startdata 
52972-1213-507 & K Nov07 & 2454437.06233 & $-$176.6 \\
 & K Nov07 & 2454437.07710 & $-$177.3 \\
53327-2044-515 & K Nov07 & 2454435.94873 & $-$193.7 \\
 & K Nov07 & 2454436.81393 & $-$193.1 \\
 & K Nov07 & 2454436.85729 & $-$193.7 \\
53436-1996-093 & K Nov07 & 2454437.09230 & $-$11.0 \\
 & K Nov07 & 2454437.10709 & $-$15.6 \\
 & K Nov07 & 2454437.13194 & $-$16.4 \\
 & K Nov07 & 2454437.14671 & $-$18.0 \\
 & K Mar08 & 2454552.81982 & $-$17.7 \\
\enddata 
\tablenotetext{a}{K = Keck, M = Magellan, V = VLT}
\tablenotetext{b}{Heliocentric radial velocity}
\tablerefs{
Note. Table 8 is published in its entirety in the electronic edition of The Astrophysical Journal. A portion is shown here for guidance regarding its form and content.
}
\end{deluxetable}

\section{TEMPERATURE AND GRAVITY DETERMINATIONS}

In order to determine the stellar atmospheric parameters necessary for
our subsequent chemical abundance analysis we have adopted three
independent techniques.  First, we employed the fitting of model
atmosphere fluxes to spectrophotometric observations.  In principal,
this provides unique values of the {\teff}, {\logg}, and metallicity
[M/H], although uncertainties in interstellar reddening, flux
calibration, and model atmosphere fluxes can influence the fits.  The
metallicity determined from the global flux is the least precisely
determined parameter, but the imprecision little affects fitting the
other two parameters.  Interstellar reddening is very important,
because it affects the fitted value of both {\teff} and {\logg}.  In
some cases, the reddening can be fitted together with the stellar
parameters, by comparing the size of the residuals between the
different reddening-corrected observations and the fitted fluxes.  In
other cases, its value can be achieved by examining the variation of
the residuals with wavelength.  Theoretical isochrones for halo stars
can also be used to constrain combinations of {\teff}, {\logg}, [M/H],
and $E(B-V)$.

A second important determinant of {\teff} is the hydrogen line
profiles, in particular those of H${\alpha}$, H${\beta}$, and
H${\gamma}$.  For {\teff} $ < $ 7000 K, the hydrogen line profile
wings are only slightly sensitive to \logg, but given an estimate of
the appropriate {\logg} from the flux fitting, an excellent
reddening-independent temperature can be derived
\citep[e.g.,][]{Fuhr93,Bark02,Aspl06,Bark08}. Care must be taken to
ensure that the continuum of the echelle spectra over the hydrogen
lines is correctly defined, but this can be done reliably using a
smooth-spectrum star or the shape of the continuum in neighbouring
orders. The treatment of convection in the 1D model atmospheres does
alter the hydrogen line profiles
\citep[e.g.,][]{Fuhr93,Bark02,Heit02,Bark08}, but these differences are
minimal for H${\alpha}$. More recently, the effects on hydrogen lines
of 3D hydrodynamical model atmospheres in self-consistently-computed
convective energy transport has been explored \citep[e.g.,][]{Ludw09},
but a systematic study of the observed spectra of 3D models still
remains to be done. In the present work we have adopted the 1D
formalism, and intend to return to this important issue in a later
study.

Finally, we also used an H$\delta$ line index (HP2) measured from
medium-dispersion spectra and calibrated as a function of {\teff} for
a large sample of stars (see Appendix). This is a very useful
technique, as it is independent of reddening, and can be used when a
flux-calibrated spectrum is unavailable.

We discuss each of these in turn.

\subsection{Spectrophotometry}\label{Sec:spectrophotometry}

The flux spectrum of a star can be considered as reflecting the
underlying blackbody temperature, moderated by the photospheric
opacity sources. In the UV-optical region for A--K stars, there are
two main continuum opacity sources: bound-free neutral hydrogen (b-f
HI) and the bound-free negative hydrogen ion (photoelectric
ionization) (b-f H$^{-}$) (see, e.g., \citealp{Gray92}). In the UV
below 3646\,\AA, absorption from HI atoms in level $n = 2$ produces
the Balmer continuum. In the optical between 3646\,\AA\ and 8206\,\AA,
absorption from HI atoms in level $n = 3$ produces the Paschen
continuum. Plotted against wavelength, the shape of the b-f HI opacity
is a series of ramps that terminate abruptly at wavelengths
corresponding to the different excitation levels of the hydrogen
atom. In contrast, the b-f H$^{-}$ opacity is smooth and approximately
bell-shaped (FWHM 10,000\,\AA), with maximum absorption at about
8500\,\AA. The b-f HI opacity dominates in A stars, while H$^{-}$
dominates for temperatures cooler than the Sun. As the ratio of the
number of H$^{-}$ to HI is proportional to the electron pressure (or
effective gravity), higher gravity increases the contribution of the
H$^{-}$ opacity relative to that from HI.

Figure~\ref{Fig:fluxes} shows the spectra of three MARCS models with
the same {\teff} and metallicity but different values of
\logg\ corresponding to the halo main sequence, subgiant branch, and
horizontal branch.  From these spectra one can see how decreasing
gravity increases the Balmer Jump (the difference in flux between the
Balmer and Paschen continua), and more subtly, makes the slope of the
Paschen continuum bluer. It is the interplay between the different
contributions made by the H$^{-}$ and HI opacity for different
temperatures and gravities that enables the gravity to be derived (for
4500K $\lesssim$ \teff\ $\lesssim$ 9000K) from fitting the
flux-calibrated spectrum \citep[e.g.,][]{Barb39, Oke65,Bess07}.

\begin{figure}
\figurenum{4}
\epsscale{1.2}
\plotone{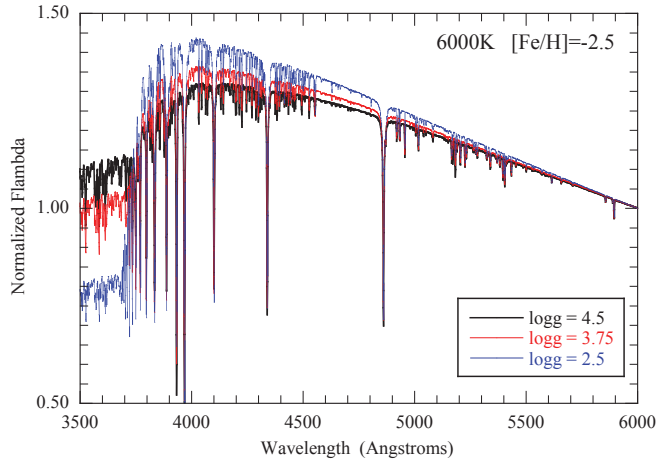}
\caption{\label{Fig:fluxes}The spectrum of three MARCS models with the
  same \teff\ = 6000\,K and metallicity [Fe/H] = $-2.5$ but different
  values of \logg\ corresponding to the main sequence, subgiant
  branch, and horizontal branch for halo stars. Note not only the
  different Balmer Jumps, but the different slopes redward of the
  Balmer Jump.}
\end{figure}

In fitting the flux spectrum of an F--K star, one should note that the
temperature is determined mainly from the slope of the Paschen
continuum; however, the strength of the hydrogen lines, which are
mainly sensitive to temperature in this spectral-type range, can serve
as valuable consistency checks on the adopted reddening and the
Paschen continuum slope fit. At medium resolution, the metallicity can
also be estimated from individual strong lines, such as Ca II and Mg
I, as well as from general metal-line blanketing in the violet, and
from the strength of molecular bands, such as CH (for carbon-normal
stars), MgH, and TiO. How well all these features are fitted is
quantitatively evaluated to determine the
spectrophotometrically-derived {\teff}, {\logg}, [M/H], and $E(B-V)$.
The precision of the spectrophotometrically-derived metallicity
differs with \teff, being higher for K stars than for F stars, but in
our experience is normally within $\pm$0.2 dex of the high-resolution
spectroscopic estimate. This is more than adequate for discovery
programs and for determining \teff\ and \logg\ values.

\subsubsection{Spectrophotometric Observations}

Medium-resolution spectra of our program stars were taken with the
ANU's 2.3m Telescope on Siding Spring Mountain, primarily with the
Double Beam Spectrograph (DBS) at 4\,\AA\ resolution, together with
some at 2\,\AA\ resolution with the Wide-Field Integral Field
Spectrograph (WiFeS). Both are double-beam spectrographs that use a
dichroic mirror to separate the blue (3000--6200\, {\AA}) and red
(6000-9700\, {\AA}) regions.  The spectra were taken using a 2
arcsec slit, with the spectrograph orientation set so the atmospheric
dispersion was along the slit. Spectrophotometric standards were also
observed each night, together with a smooth-spectrum star to enable
the removal of telluric features \citep[e.g.,][]{Bess99}.  We used an
updated list of 
standards\footnote{http://www.mso.anu.edu.au/$\sim$bessell/FTP/Spectrophotometry/}
that were selected from the Next Generation Spectral Library (NGSL) 
\citep{Heap07}\footnote{http://archive.stsci.edu/prepds/stisngsl/index.html}. 

The CCD frames were bias subtracted and flat fielded, the cosmic rays
removed, the star and sky spectra extracted, and the sky removed. All
spectra were divided by the extracted spectrum of a smooth-spectrum
star (generally EG131 or L745-46a, cool He white dwarfs) to remove the
signatures of the grating and CCD response, and in the red, any
telluric absorption, resulting in a gently curved spectrum. The
spectra were then wavelength fitted, wavelength scrunched (rebinned to
a linear scale), and corrected for the continuous atmospheric
extinction appropriate for their observed airmass. The standard-star
spectra (generally 8--10 per night) were compared with their standard
values, and the mean spectrophotometric calibration determined and
applied to all the program stars.

\subsubsection{Model Atmosphere Fluxes}

We initially used the
\citet{Muna05}\footnote{http://archives.pd.astro.it/2500-10500/} 
library of synthetic spectra at 1\,\AA\ resolution, but have recently
added the LTE model atmosphere fluxes from the MARCS grid
\citep{Gust08}\footnote{http://marcs.astro.uu.se}. The MARCS spectra
are not line-by-line computed spectra (as are the Munari et
al. spectra), but are fluxes generated using statistically sampled
opacities; however, smoothed to 4\,\AA\ resolution they are very good
representations of real spectra. Although restricted to temperatures
below 8000\,K, the MARCS model atmosphere grid covers all the
parameter space of the halo stars that we are interested in.

The \citet{Muna05} spectra cover a wide range of atmospheric
parameters: 3500\,K $<$ \teff\ $<$ 47500\,K, 0.0 $<$ \logg\ $<$ 5.0,
and $-2.5$ $<$ [M/H] $<$ +0.5.  Extension of the full grid to lower
metallicity is underway by R. Sordo (2010, private communication), but
many relevant lower metallicity models are already available\footnote{http://www.user.oat.ts.astro.it/castelli/spectra.html}. 

For the MARCS spectra, the stellar atmospheric model parameters ranged
in \teff\ from 2500\,K to 8000\,K, in steps of 100\,K from 2500\,K to
4000\,K, and in steps of 250\,K between 4000\,K and 8000\,K. The
\logg\ values were between $-1.0$ and 5.5 in steps of 0.5. Overall
logarithmic metallicities relative to the Sun were between $-5.0$ and
$+1.0$ in variable steps. The reference solar abundance mixture was
that of \citet{Grev07}. Plane-parallel models were used for gravities
between 3.0 and 5.0, and spherical models (1M$_{\odot}$) for lower
gravities. For the lower-metallicity stars, alpha-enhanced models were
used: [$\alpha$/Fe] = +0.25, for $-0.5\leq$ [Fe/H] $\leq -1.5$, and
+0.50, for $-1.5<$ [Fe/H] $\leq -5.0$.

\subsubsection{Spectrophotometric Flux Fitting Method}

A \textit{python}\footnote{http://www.python.org/psf} program,
\textit{fitter}, written by S.\ J.\ M., was used for the fitting. This
will be described in greater detail in a later paper, together with
parameters for many more stars. Basically, the grid of fluxes was
initially interpolated to produce a new grid at spacings of 100\,K in
\teff, 0.5 in {\logg}, and 0.1 in [M/H]. The software was run using
parallel-processing architecture, with each of the 25 processors
allocated model spectra for a single metallicity.

The model fluxes were smoothed to a resolution similar to that of the
observations and the observed fluxes and the model fluxes renormalized
to the mean flux between 4500 and 5500\,\AA. The radial velocity of
the observed star was obtained through cross-correlation of deep lines
with a model spectrum, and the observed spectrum shifted to match the
model. We were also able to mask out regions of the observed spectrum
that we wish to ignore in the fitting, such as the regions around the
CH and C$_{2}$ bands in C-rich stars. Each model spectrum in the new
grid was then cross-correlated against the observed spectrum, the RMS
of the fit computed and the parameters of the best-fitting grid
spectrum identified.  A new grid centered on these parameters was then
interpolated at a finer spacing of 25\,K in \teff\ and 0.1~dex in
\logg, and these finer-spaced grid spectra were cross-correlated
against the observed spectrum, the RMS of the fit calculated, and the
best-fitting parameters again selected. Fitting only the blue spectrum
was found to be adequate for most halo stars, but we did fit the
combined blue and red spectrum from 3600\,\AA\ to 9000\,\AA.
Figure~\ref{Fig:fitter} shows the result of one such fit for the blue
spectrum of the metal-poor star HE~1311-0131, using the MARCS
grid\footnote{We have not used the wavelength region
  3700-3900\,{\AA}, which includes the confluence of the Balmer
  series, since experience shows this detracts from the goodness of
  fit we obtain for the Balmer discontinuity.  When spectra of
  carbon-rich stars with obviously strong CH and/or C$_{2}$ bands were
  fitted with scaled solar metallicity model atmosphere fluxes, the CH
  and C$_{2}$ bands were masked out in the fit.}.

\begin{figure*}
\figurenum{5}
\plotone{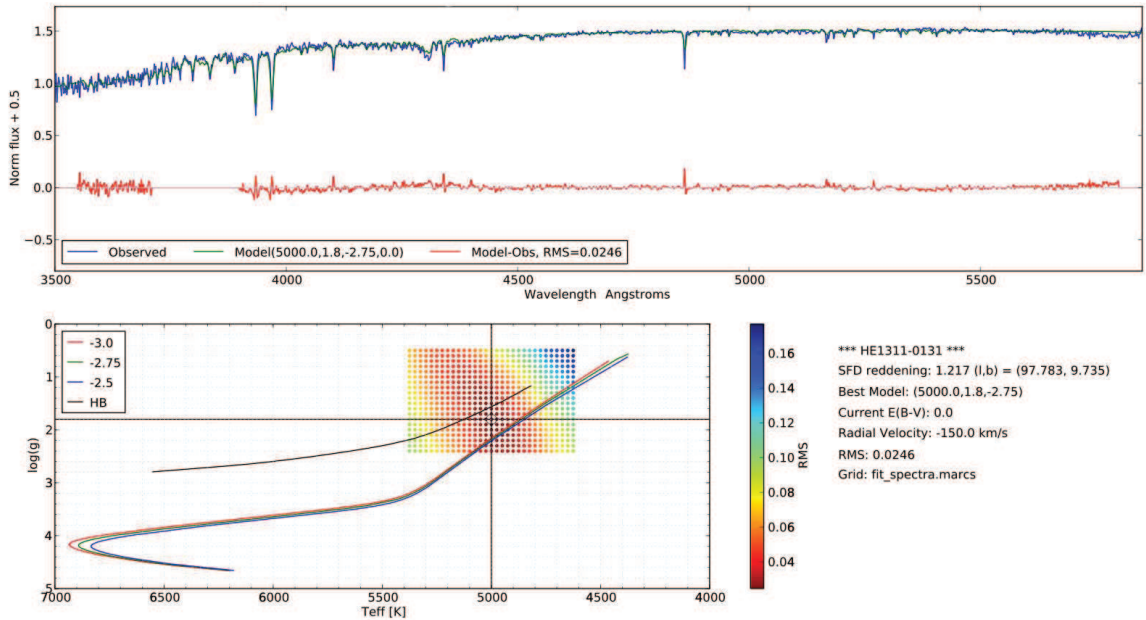}
\caption{\label{Fig:fitter}The output of \textit{fitter}. In the top
  frame the blue line represents the observed spectrum, the green line
  is the best fitting model spectrum, and the red line is the
  difference spectrum. In the lower frame the cross-hair marks the
  best fitting \teff\ and \logg\ and the RMS values of the fits are
  represented in color for the small subgrid about the initial best
  fitting parameters. The RMS-color palette is shown ranging from
  black for best to blue for worst. Any combination of parameters
  within the black shaded region is acceptable. The halo isochrones
  are drawn to assist the selection of the most likely parameters.}
\end{figure*}

Revised fits of fluxes corrected for interstellar reddening
corresponding to a single value, or a range of $E(B-V)$ values, can
also be assessed. The reddening curve used is from \citet{Math90}. To
assist in estimating the reddening, the \citet{schlegel98} reddening
(maximum possible value in that direction) is indicated on the output
for each star. It was also noted that there is strong evidence for a
reddening-free bubble within 100\,pc of the Sun
\citep[e.g.,][]{Frisch11,Abt11} that likely extends even farther in the
directions of the Galactic poles.  It should also be kept in mind that
the continuum temperatures derived with no reddening are minimum
temperatures, as any corrections for reddening will increase the
fitted temperatures.  The spectrophotometrically-derived {\teff},
\logg, and $E(B-V)$ are given in columns (2), (3), and (4) of
Table~\ref{Tab:teff}. We choose not to list the derived [M/H] value, but
defer to the precisely determined values in Paper II.

\subsection{H$\alpha$, H$\beta$, and H$\gamma$ Line Profile Fitting}

In F, G, and K stars, where the negative hydrogen ion is the dominant
source of continuous opacity, hydrogen Balmer-line wings are extremely
useful indicators of \teff, since their strength is, to a large
degree, independent of other parameters such as hydrogen abundance,
gravity, and chemical composition \citep[e.g.,][]{Fuhr93}.
\citet{Bark08} discusses the pros and cons in the use of hydrogen
Balmer lines as high-precision diagnostics of effective temperature in
these stars, and highlights the improvements in detectors and
broadening theory that underpin this. \citet{Bark03} detail publicly
available codes that can be used to compute Balmer-line profiles, and
include the advances in broadening theory.  However, \citet{Bark08}
notes the role of possible departures from LTE, differences between 1D
and 3D model atmospheres, and the effects that the value of the
mixing-length parameter has on the Balmer-line profiles.

The \teff\ were derived by P.S.B. from fitting the H$\alpha$,
H$\beta$, and H$\gamma$ profiles measured from our echelle
spectra\footnote{We chose not to analyze H$\delta$ because of its
  lower S/N on our spectra, which affects both local fitting and the
  ability to reliably apply the continuum normalisation techniques
  used here.} described in Section~\ref{Sec:spectra}.  The method
employed for merging of spectral orders and continuum normalisation of
the spectra and subsequent analysis follows precisely that described
in \citet{Bark02}.  The most important aspects of the analysis are as
follows.  Synthetic profiles are computed assuming LTE line formation
using one-dimensional LTE plane-parallel MARCS models \citep{Aspl97},
with convection described by mixing-length theory with parameters
$\alpha = 0.5$ and $y = 0.5$. The most important line-broadening
mechanisms for the wings are Stark broadening and self broadening,
which are described by calculations of \citet{Stehle99} and
\citet{Bark00}, respectively. The fitting is done by minimization of
the $\chi^2$ statistic, comparing the observed and synthetic profiles
in spectral windows believed to be free of blends in the solar
spectrum.  To illustrate the fitting method and the temperature
sensitivity of the hydrogen line profiles, Figure~\ref{Fig:hydrogen}
shows fits to H$\alpha$, H$\beta$, and H$\gamma$ line profiles for
HE~2047$-$5612, a typical halo subgiant.

\begin{figure*}
\figurenum{6}
\epsscale{0.8}
\plotone{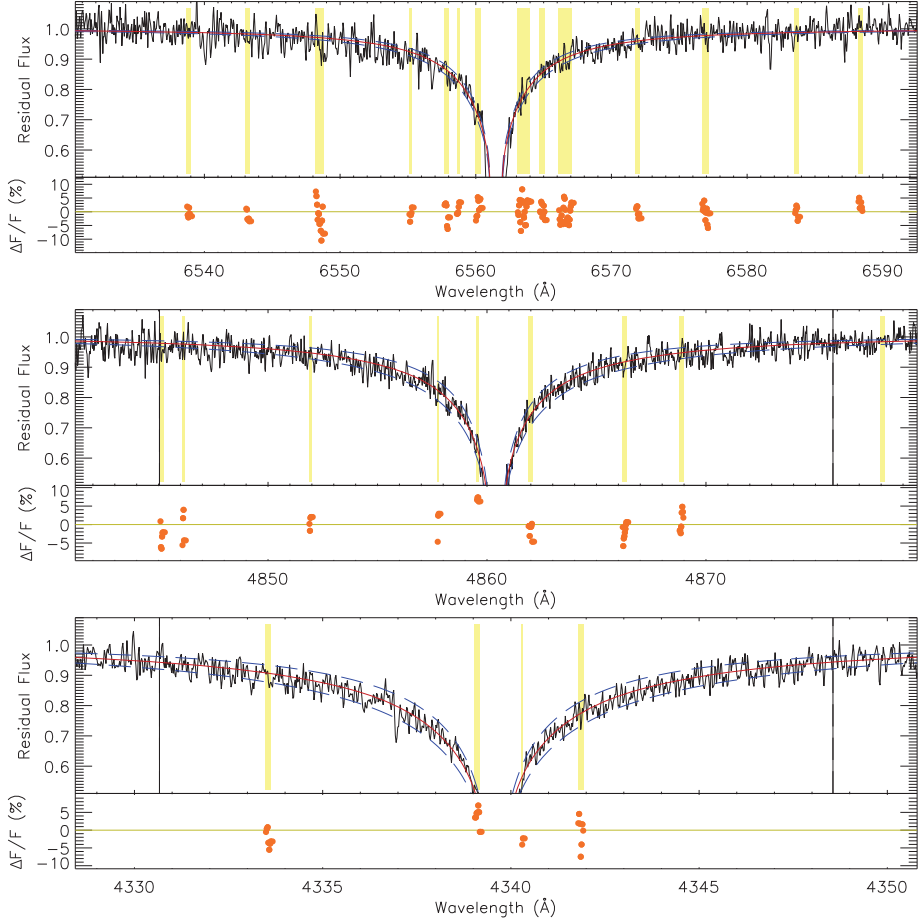}
\caption{\label{Fig:hydrogen}Fits to H$\alpha$, H$\beta$, and
  H$\gamma$ line profiles for HE~2047$-$5612, \teff$\mathrm{(Balmer)}
  = 6040$~K, assuming $\log g = 3.6$ and [Fe/H]$=-3.2$.  The (black)
  full lines showing noise are the normalised observed spectra and the
  (red) smooth full lines the best fit synthetic profile
  (corresponding to \teff$ = $6040, 6020, 6080~K for each line,
  respectively).  The (blue) dashed lines show the synthetic spectra
  calculated with models 200~K cooler and hotter than the best
  fit. The (yellow) shaded regions show the windows used for
  determining the $\chi^2$ statistic in the fitting, and residuals for
  these regions are shown in the bottom panel.  The full vertical
  lines show the estimated limit of validity of the impact
  approximation in the self-broadening calculations, which is beyond
  the limit of the plot for H$\alpha$. Note that the windows outside
  this region are rejected in the fitting, and thus no residual is
  plotted. }
\end{figure*}

The choice to use LTE analysis follows the reasoning presented in
\citet{Aoki09} and bears repeating here.  The assumption of LTE for
the line wings has been shown to be questionable on the basis of
theoretical non-LTE calculations \citep{Bark07}, the role of hydrogen
collisions being a major uncertainty.  Those calculations suggest that
the temperatures from LTE Balmer-line wings could be systematically
too cool, by of order 100\,K if hydrogen collisions are inefficient,
although if collisions are efficient, LTE is not ruled out. Since
there is no strong evidence favoring any particular hydrogen collision
model, we calculate in LTE, as this temperature scale is well studied,
and it is computationally most practical.  However, we emphasize that
LTE is not a safe middle ground, and will lead to temperatures that
are systematically too cool, should departures from LTE exist in
reality.

The individual Balmer lines have distinct characteristics and
behaviors; so, as in \citet{Aoki09}, we chose not to give \teff\ from
all lines equal weight in determining our final Balmer-line
\teff. H$\alpha$ is often preferred over other lines in solar-type
stars, for the reasons discussed by \citet{Fuhr93}.  However, in
metal-poor stars it is not clear that H$\alpha$ is to be preferred.
Blending by metal lines becomes unimportant, and H$\beta$ becomes
insensitive to gravity, while H$\alpha$ is quite gravity sensitive
(see \citealp[their Table~4]{Bark02}) and rather insensitive to
\teff\ (see Figure~\ref{Fig:hydrogen}).  These differences in behavior
arise due to changes in the relative importance of Stark and self
broadening.  Moreover, the calculations by \citet{Bark07} suggest that
non-LTE effects, if they exist, will be largest in
H$\alpha$. H$\gamma$, while having similar sensitivities to H$\beta$,
is generally not as reliable as H$\beta$.  This is predominantly due
to increased blending in this region of the spectrum, especially in
CEMP stars with strong G bands, which affects both the local fitting
of the line and nearby orders used to define the continuum placement.
In addition, the $S/N$ is generally much lower at H$\gamma$ due to
less flux in the blue.  Considering all these factors, we judge
H$\beta$ as the most reliable line for determining {\teff}, and so in
combining the temperatures from H$\alpha$, H$\beta$, and H$\gamma$, we
have given H$\beta$ double weight.

Thus, having determined \teff\ from each line for each spectrum
(\teff(H$\alpha$), \teff(H$\beta$), \teff(H$\gamma$)), to determine
the final hydrogen line profile \teff(Balmer) for a star we employed
the following procedure.  For stars with multiple spectra, we first
determined the mean \teff(H$\alpha$), \teff(H$\beta$),
\teff(H$\gamma$) for the star, weighting each determination by the
$S/N$ of the spectra. Next, we combined the mean \teff(H$\alpha$),
mean \teff(H$\beta$), and mean \teff(H$\gamma$) temperatures using the
1:2:1 weighting into a final average \teff(Balmer).  Because the
profiles are slightly dependent on the value of \logg, where the
gravity adopted for the hydrogen line fits differed grossly from that
derived from fitting the blue fluxes, the spectrophotometric gravity
was used and the hydrogen lines were refitted to obtain a second
temperature estimate.  In such cases, where multiple analyses were
performed with different \logg, the standard deviation of the mean
\teff(H$\alpha$), \teff(H$\beta$), and \teff(H$\gamma$) values was
calculated, and we selected \teff(Balmer) as the one with the smallest
dispersion.  Conversely, where the temperature derived from the
hydrogen lines grossly exceeded the continuum-fitted temperature, the
application of additional reddening was tested to see whether a higher
temperature would give an acceptable fit.  The final hydrogen line
profile temperatures, \teff(Balmer), are given in column (5) of
Table~\ref{Tab:teff}, together with the standard deviation,
$\sigma$({\teff}), of the different hydrogen lines (column (6)) and the
number of individual hydrogen line observations used (column (7)).
The adopted gravity is given in column (8).

\subsection{Medium Resolution H$\delta$ Index HP2}

A third, independent, temperature estimate was obtained from
medium-resolution spectroscopy of the H$\delta$ line index (HP2),
calibrated as a function of {\teff} as described in the Appendix.  We
determined HP2 from the medium-resolution spectra of the candidate
most metal-poor stars we observed with the 2.3m/DBS combination, and
those we obtained from the SDSS Data Release 7, as described in
Section~\ref{Sec:discovery}.  We now use the above calibration to estimate
their effective temperatures, excluding two categories of objects.
First, while our calibration may be applied with confidence for
``normal'' metal-poor stars, it would be inappropriate to use it for
carbon-rich objects, because of CH absorption in the bandpasses used
to measure HP2.  We thus do not present temperatures for the seven
objects in our program sample that exhibit strong G~bands:
specifically, we exclude C-rich stars for which the \citet{beers99}
G-band index G$^{\prime}$ is larger than $\sim1.0$\,{\AA} (see
Table~\ref{Tab:sample}).  Secondly, we also exclude stars with fewer
than 400 counts/1\,{\AA} pixel at 4100\,{\AA}, in an effort to exclude
objects for which the poorer S/N might be expected to decrease the
accuracy of the temperature estimate.  The values of HP2 and the
inferred values of {\teff} for the remaining 26 C-normal stars in our
most metal-poor star sample are presented in columns (9) and (10) of
Table~\ref{Tab:teff}.

\subsection{Possible Systematic Differences in Temperature between Techniques}

Figure~\ref{Fig:deltat} plots the differences between the temperature
derived from fitting the overall flux and the temperatures derived
from high-resolution hydrogen line profile fitting (LHS) and the
H$\delta$ index (RHS).  There are indications of some systematic
differences in these figures. Compared to the temperatures derived
from fitting the fluxes, the hydrogen line profiles yielded lower
temperatures by 128 $\pm$ 18\,K (36 stars), while the H$\delta$ index
yielded temperatures lower by 8 $\pm$ 22\,K (25 stars).

\begin{figure}
\figurenum{7}
\epsscale{1.2} 
\plotone{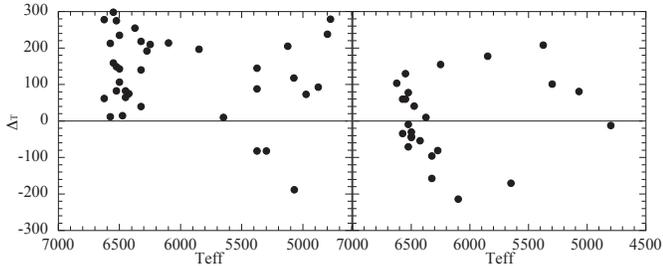}
\caption{\label{Fig:deltat}Differences between the \teff\ derived from
  flux fitting and those derived from, H$\alpha$, H$\beta$, and
  H$\gamma$ profiles (left), and the medium-resolution H$\delta$
  index, HP2 (right).}
\end{figure}

As mentioned earlier, the temperatures derived from the hydrogen line
profiles may be affected by the neglect of non-LTE and by remaining
uncertainties in the broadening theory.  The temperatures from flux
fitting and hydrogen line fitting may both be affected by the use of
1D model atmospheres rather than 3D model atmospheres, but 3D effects
are more likely for the hydrogen lines than the stellar energy
distributions \citep{Aspl05}. We also note that the temperatures from
flux fitting are in agreement with those from the infrared flux method
for those metal-poor stars in common. Future work by P.S.B. and
M.A. is proposed to ascertain the reason or reasons for the cooler
hydrogen line temperatures. There is an insignificant difference
between the mean flux temperature and the empirically-calibrated mean
H$\delta$ index temperature.

In the absence of compelling reasons for discounting any of the
temperature derivation techniques, we have taken a mean of the derived
temperatures to use for our high resolution abundance analysis (Paper
II), but note that these mean temperatures may be $\approx$ 50\,K
cooler than the infrared flux method temperature scale.  Our final
\teff\ and their standard error of the mean are presented in columns
(11) and (12) of Table~\ref{Tab:teff}.

\begin{deluxetable*}{lrrrrrrrrrrr}
\tabletypesize{\scriptsize}
\tablenum{9}
\tablecolumns{12} 
\tablewidth{0pc} 
\tablecaption{\label{Tab:teff}{\teff}, LOG G, $E(B-V)$, AND HP2 FOR PROGRAM STARS}
\tablehead{ 
\colhead{Star} & 
\colhead{{\teff}\tablenotemark{a}} & 
\colhead{{\logg}\tablenotemark{a}} &
\colhead{{$E(B-V)$}\tablenotemark{a}} &
\colhead{{\teff}\tablenotemark{b}} & 
\colhead{$\sigma$({\teff})\tablenotemark{b}} &
\colhead{No.\ of \tablenotemark{b}} &
\colhead{{\logg}\tablenotemark{b}} &
\colhead{HP2}       & 
\colhead{{\teff}\tablenotemark{c}} &
\colhead{{\teff}} &
\colhead{s.e.({\teff})} 
\\
\colhead{} & 
\colhead{(K)} & 
\colhead{} &
\colhead{mag} &
\colhead{(K)} & 
\colhead{(K)} & 
\colhead{H lines} & 
\colhead{} & 
\colhead{({\AA})} &
\colhead{(K)} &
\colhead{(K)} &
\colhead{(K)} 
\\ 
\colhead{(1)} & 
\colhead{(2)} & 
\colhead{(3)} & 
\colhead{(4)} &
\colhead{(5)} &
\colhead{(6)} &
\colhead{(7)} &
\colhead{(8)} &
\colhead{(9)} &
\colhead{(10)} &
\colhead{(11)} &
\colhead{(12)} 
}

\startdata 
52972-1213-507\tablenotemark{d} &  ... &  ... &  ... & 6463 & 149 &  2 & 4.50 &   ...  &  ... & 6463 & ... \\
53327-2044-515                  & 5650 & 3.50 & 0.14 & 5640 &  85 &  2 & 4.50 &  2.74  & 5820 & 5703 &  58 \\
53436-1996-093                  &  ... &  ... &  ... & 6442 &  12 &  5 & 4.30 &  4.57  & 6455 & 6449 &   6 \\
54142-2667-094                  & 6475 & 4.10 & 0.03 & 6460 &  21 &  2 & 3.90 &  4.48  & 6434 & 6456 &  11 \\
BS~16545-089                    & 6625 & 4.00 & 0.00 & 6347 &  40 &  3 & 3.80 &   ...  &  ... & 6486 & 139 \\
CS~30336-049                    & 4800 & 2.40 & 0.05 & 4562 &  67 &  3 & 1.50 &  1.03  & 4812 & 4725 &  81 \\
HE~0049--3948                   & 6500 & 4.10 & 0.00 & 6357 &  75 &  3 & 3.80 &  5.02  & 6542 & 6466 &  55 \\
HE~0057--5959                   & 5375 & 3.30 & 0.00 & 5230 &  61 &  3 & 2.40 &  1.56  & 5167 & 5257 &  61 \\
HE~0102--1213                   & 6100 & 3.80 & 0.03 & 5886 &  35 &  2 & 3.80 &  4.03  & 6314 & 6100 & 123 \\
HE~0146--1548\tablenotemark{d}  & 4775 & 1.70 & 0.00 & 4496 &  57 &  2 & 1.20 &   ...  &  ... & 4636 & 139 \\
HE~0207--1423\tablenotemark{d}  & 5125 & 2.40 & 0.00 & 4920 &   0 &  1 & 1.80 &   ...  &  ... & 5023 & 102 \\
HE~0228--4047                   & 6575 & 4.30 & 0.02 & 6362 &  87 &  3 & 3.70 &  5.55  & 6609 & 6515 &  77 \\
HE~0231--6025                   & 6500 & 4.10 & 0.00 & 6265 &  35 &  3 & 4.20 &  5.04  & 6545 & 6437 &  86 \\
HE~0253--1331                   & 6500 & 4.10 & 0.02 & 6393 &   7 &  2 & 4.50 &  4.95  & 6530 & 6474 &  41 \\
HE~0314--1739                   & 6625 & 4.40 & 0.04 & 6563 &   7 &  2 & 4.30 &  4.90  & 6521 & 6570 &  30 \\
HE~0355--3728                   & 6450 & 4.20 & 0.01 & 6385 &  35 &  3 & 3.90 &   ...  &  ... & 6418 &  32 \\
HE~0945--1435                   & 6325 & 4.40 & 0.04 & 6285 &  53 &  3 & 3.60 &  4.43  & 6421 & 6344 &  40 \\
HE~1055+0104                    & 6375 & 4.30 & 0.03 & 6120 &  42 &  2 & 3.80 &  4.21  & 6365 & 6287 &  83 \\
HE~1116--0054                   & 6550 & 4.30 & 0.07 & 6391 &  20 & 11 & 3.70 &  4.43  & 6420 & 6454 &  48 \\
HE~1142--1422                   & 6275 & 2.80 & 0.06 & 6083 &  71 &  3 & 2.60 &  4.18  & 6356 & 6238 &  80 \\
HE~1201--1512                   & 5850 & 4.20 & 0.03 & 5653 & 125 &  6 & 4.50 &  2.45  & 5672 & 5725 &  62 \\
HE~1204--0744                   & 6525 & 4.10 & 0.05 & 6442 &  87 &  3 & 4.40 &  5.01  & 6534 & 6500 &  29 \\
HE~1207--3108                   & 5300 & 2.60 & 0.01 & 5382 &  60 &  3 & 3.10 &  1.61  & 5199 & 5294 &  52 \\
HE~1320--2952                   & 5070 & 2.40 & 0.04 & 5258 &  44 &  6 & 2.00 &  1.29  & 4989 & 5106 &  79 \\
HE~1346--0427                   & 6325 & 4.40 & 0.00 & 6185 &  81 &  3 & 3.80 &   ...  &  ... & 6255 &  70 \\
HE~1402--0523                   & 6425 & 4.20 & 0.02 & 6350 &  38 &  3 & 3.80 &  4.68  & 6479 & 6418 &  37 \\
HE~1506--0113\tablenotemark{d}  & 5075 & 2.40 & 0.05 & 4957 &  40 &  3 & 2.20 &   ...  &  ... & 5016 &  59 \\
HE~2020--5228                   & 6325 & 4.00 & 0.03 & 6107 &  40 &  3 & 3.80 &  4.70  & 6482 & 6305 & 108 \\
HE~2032--5633                   & 6525 & 4.10 & 0.05 & 6250 & 180 &  3 & 3.70 &  5.42  & 6596 & 6457 & 105 \\
HE~2047--5612                   & 6250 & 4.00 & 0.05 & 6040 &  31 &  3 & 3.60 &  3.39  & 6095 & 6128 &  62 \\
HE~2135--1924                   & 6525 & 4.40 & 0.03 & 6376 &  57 &  2 & 4.50 &  4.54  & 6447 & 6449 &  43 \\
HE~2136--6030                   & 6450 & 4.00 & 0.00 & 6367 & 123 &  3 & 3.80 &   ...  &  ... & 6409 &  41 \\
HE~2139--5432\tablenotemark{d}  & 5375 & 2.20 & 0.01 & 5457 &  44 &  3 & 2.00 &   ...  &  ... & 5416 &  41 \\
HE~2141--0726                   & 6575 & 4.40 & 0.03 & 6563 &  78 &  2 & 4.20 &  4.86  & 6515 & 6551 &  18 \\
HE~2142--5656\tablenotemark{d}  & 4975 & 2.00 & 0.04 & 4902 & 139 &  3 & 1.60 &   ...  &  ... & 4939 &  36 \\
HE~2202--4831\tablenotemark{d}  & 5375 & 2.40 & 0.00 & 5287 &  85 &  3 & 2.20 &   ...  &  ... & 5331 &  44 \\
HE~2246--2410                   & 6550 & 4.20 & 0.01 & 6252 & 164 &  3 & 4.50 &  4.74  & 6490 & 6431 &  90 \\
HE~2247--7400\tablenotemark{d}  & 4875 & 2.00 & 0.01 & 4782 &  40 &  3 & 1.60 &   ...  &  ... & 4829 &  46 \\
\enddata 

\tablenotetext{a}{From fit to spectrophotometric flux.}
\tablenotetext{b}{From fit to H$\alpha$, H$\beta$, and H$\gamma$ profiles.}
\tablenotetext{c}{From empirically calibrated H$\delta$ index (HP2). See Appendix.}
\tablenotetext{d}{C-rich star.}
\end{deluxetable*}

While spectrophotometry, in principle, remains the best way of
obtaining stellar parameters, modern wide-field broad-band multicolor
surveys such as SDSS, and in particular SkyMapper \citep{Keller07},
will provide precise photometric indices that can be calibrated using
synthetic photometry from model atmosphere fluxes to provide accurate
temperatures, and good estimates of gravity and
metallicity. Interstellar reddening, however, remains an issue for all
photometric techniques, emphasizing the continuing necessity of
hydrogen line fitting and the importance of further theoretical work
on the formation of hydrogen lines in cool stars.

\section{SUMMARY}

We report the discovery of 34 stars in the Hamburg/ESO Survey for
metal-poor stars and the Sloan Digital Sky Survey that have [Fe/H]
$\la$ --3.0.  Ten of them are newly discovered objects having [Fe/H]
$<$ --3.5.  We have obtained high-resolution, high-$S/N$ spectra of
them and four other extremely metal-poor stars (three of which have
[Fe/H] $<$ --3.5), and present equivalent widths and radial velocities
for this sample.

\teff\ has been determined for these objects, employing three
independent techniques.  First, we analyzed medium-resolution spectra
to obtain absolute fluxes. These were fit using model atmosphere
fluxes to provide spectrophotometric {\teff} and \logg, with
approximate [M/H] and reddening.  Second, we fit the wings of the
H$\alpha$, H$\beta$, and H$\gamma$ lines, measured from our echelle
spectra, to model atmosphere line profiles.  Although there are some
caveats concerning our understanding of the formation of these lines,
this technique provides a reliable and reddening-independent method of
temperature determination.  Finally, we used the observed H$\delta$
index, HP2, together with a calibration of HP2 as a function of \teff,
for a set of stars that have well-established temperatures, to obtain
a third \teff\ estimate.

There are possible systematic differences in the temperatures derived
using these three techniques, of order 100\,K, and future work is
needed to clarify the origin of these differences.  Nevertheless, for
our purposes, we adopted the average of the three determinations.  The
mean (internal) error of the resulting temperatures is 63\, K, with a
dispersion of 33\, K.

The data presented here have been analyzed in Papers II, III, and IV
of this series (Yong et al. 2012a, b, and Norris et al. 2012a,
respectively), which includes a homogeneous re-analysis of similar
data available in the literature.  This has yielded relative chemical
abundances for a total of $\sim$86 stars having [Fe/H] $\la$ --3.0
(and some 32 with [Fe/H] $\la$ --3.5), which have been used to further
constrain the conditions that existed at the earliest times.

\acknowledgments

J.\ E.\ N., M.\ S.\ B., D.\ Y., and M. A. \ gratefully acknowledge
support from the Australian Research Council (grants DP03042613,
DP0663562, DP0984924, and FL110100012) for studies of the Galaxy's
most metal-poor stars and ultra-faint satellite systems.
J.\ E.\ N.\ and D.\ Y.\ acknowledge financial support from the Access
to Major Research Facilities Program, under the International
Science Linkages Program of the Australian Federal Government.
Australian access to the Magellan Telescopes was supported through the
Major National Research Facilities program.  Observations with the
Keck Telescope were made under Gemini exchange time programs
GN-2007B-C-20 and GN-2008A-C-6.
N.\ C.\ acknowledges financial support for this work through the
Global Networks program of Universit\"at Heidelberg and
Sonderforschungsbereich SFB 881 ``The Milky Way System'' (subproject
A4) of the German Research Foundation (DFG).
P.S.B acknowledges support from the Royal Swedish Academy of Sciences
and the Swedish Research Council; he is a Royal Swedish Academy of
Sciences Research Fellow supported by a grant from the Knut and Alice
Wallenberg Foundation.  
T.\ C.\ B.\ acknowledges partial funding of this work from grants PHY
02-16783 and PHY 08-22648: Physics Frontier Center/Joint Institute for
Nuclear Astrophysics (JINA), awarded by the U.S. National Science
Foundation.  
The authors wish to recognize and acknowledge the very significant
cultural role and reverence that the summit of Mauna Kea has always
had within the indigenous Hawaiian community. We are most fortunate to
have the opportunity to conduct observations from this mountain.
Finally, we are pleased to acknowledge support from the European
Southern Observatory's Director's Discretionary Time Program.
}

\appendix


\subsection{\teff\ Calibration of the H$\delta$ Index HP2}

\begin{figure}
\figurenum{A1}
\epsscale{0.5} 
\plotone{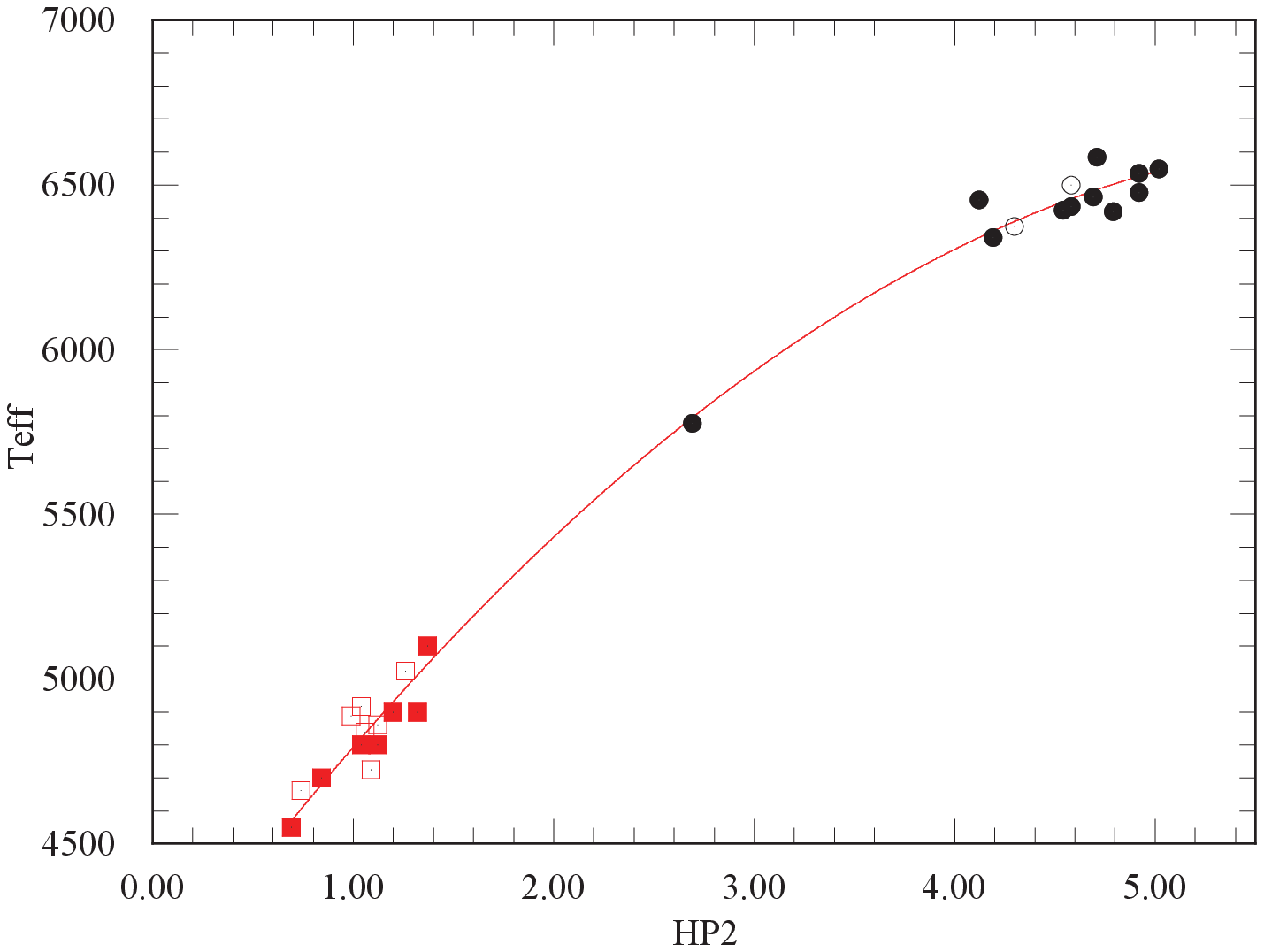}
\caption{\label{Fig:appendix}\teff\ versus HP2 for calibrating
  stars.  Filled black circles represent dwarfs (and one
    subgiant) from \citet{Casa10}; open black circles are dwarfs from
    the present work; filled red boxes are giants from
    \citet{cayrel04}; and open red boxes are the average values for
    giants from \citet{cayrel04} and the present work. The red line is
    the fitted quadratic relation.  }
\end{figure}

A very useful effective temperature estimate for metal-poor stars can
be obtained from medium-resolution spectroscopy of the H$\delta$ line
index (HP2) calibrated as a function of {\teff}, following
\citet{Ryan99}.  We used spectra of metal-poor dwarfs and giants that
covered the region around H$\delta$, obtained with ANU's 2.3\,m
Telescope/Double Beam Spectrograph (DBS) combination on Siding Spring
Mountain, as described in Section~\ref{Sec:2.3m} for our observations
of candidate most metal-poor stars.  The sample comprised 28 stars
with abundances in the range --4.0 $<$ [Fe/H] $<$ --2.5 for which
temperatures had been determined using a combination of three
independent techniques.  First, results for 12 dwarfs and subgiants
with {\teff} in the range 5800--6600\,K, have been taken from the work
of Casagrande et al. (2010), who used the infrared flux method.
Second, 14 metal-poor red giants having 4650\,K $<$ {\teff} $<$
5100\,K came from \citet[][their Table 4]{cayrel04}, who base their
temperatures on calibrations of broadband $B-V$, $V-R$, $V-K$, and
$V-I$ photometry.  Finally, results for nine dwarfs and giants were
determined by M.\ S.\ B. for the present work using the
spectrophotometric techniques described in
Section~\ref{Sec:spectrophotometry}. From the DBS spectra, we measured
the H$\delta$ index, HP2, of \citet[][their Section 3.1.1]{beers99}.
The data for our calibration stars are presented in
Table~10, where columns (1)--(4) contain
identification, HP2, {\teff}, and sources, respectively.  It is noted
here, for completeness, that for the seven red giants with two
determinations of {\teff} the absolute mean difference was 5\,K with
dispersion 56\,K.  The calibration data are plotted in
Figure~\ref{Fig:appendix} together with the fitted quadratic
relationship {\teff} = 4018. + $842.6\times$HP2 --
$67.70\times$HP2$^{2}$, about which the RMS deviation is 64\,K.

\begin{deluxetable}{lccc}
\tabletypesize{\scriptsize}
\tablenum{10}
\tablecolumns{5} 
\tablewidth{0pc} 
\tablecaption{\label{Tab:appendix}H$\delta$ INDICES AND {\teff} FOR CALIBRATION STARS}
\tablehead{ 
\colhead{Star} & \colhead{HP2} & \colhead{{\teff}} & \colhead{Source\tablenotemark{a}} \\ 
\colhead{} & \colhead{({\AA})} & \colhead{{(K)}} & \colhead{} \\  
\colhead{(1)} & \colhead{(2)} & \colhead{{(3)}} & \colhead{(4)}  
}

\startdata 

BD~+3 740    &  4.79 &  6419 & 1 \\
BD~+9 2190   &  4.92 &  6477 & 1 \\
BD~+26 3578  &  4.54 &  6425 & 1 \\
BD~--13 3442  &  4.58 &  6434 & 1 \\
CD~--24 17504 &  4.12 &  6455 & 1 \\
CD~--33 1173  &  5.02 &  6548 & 1 \\
CD~--38 245   &  1.12 &  4800 & 2 \\
BS~16477-003 &  1.32 &  4900 & 2 \\
CS~22169-035 &  0.74 &  4662 & 2,3 \\
CS~22172-002 &  1.06 &  4838 & 2,3 \\
CS~22186-025 &  1.20 &  4900 & 2 \\ 
CS~22189-009 &  0.99 &  4887 & 2,3 \\
CS~22873-166 &  0.69 &  4550 & 2 \\
CS~22885-096 &  1.26 &  5025 & 2,3 \\
CS~22891-209 &  1.09 &  4725 & 2,3 \\
CS~22897-008 &  1.04 &  4918 & 2,3 \\
CS~22952-015 &  1.04 &  4800 & 2 \\
CS~22953-003 &  1.37 &  5100 & 2 \\
CS~22968-014 &  1.12 &  4862 & 2,3 \\
CS~29491-053 &  0.84 &  4700 & 2 \\
G~4-37       &  4.19 &  6340 & 1 \\
G~64-12      &  4.69 &  6464 & 1 \\
G~64-37      &  4.71 &  6584 & 1 \\
G~186-26     &  4.30 &  6375 & 3 \\
HD~140283    &  2.69 &  5777 & 1 \\
LP~651-4     &  4.58 &  6500 & 3 \\
LP~815-43    &  4.92 &  6535 & 1 \\

\enddata 
\tablenotetext{a}{{\teff} sources: 1 = Casagrande et al. (2010); 2 = Cayrel et al. (2004); and 3 = Present work}
\end{deluxetable}

\noindent{\it Facilities:} {ATT(DBS); Keck:I(HIRES);
  Magellan:Clay(MIKE); VLT:Kueyen(UVES)}

\end{document}